\documentclass[12pt]{amsart}
\usepackage{amssymb,epsfig}
\usepackage[vcentermath]{youngtab}

\newtheorem{propo}{Proposition}[section]
\newtheorem{theo}{Theorem}[section] 

\newenvironment{demo}[1]{\textit{Proof #1. }}{\hfill$\diamondsuit$}

\newtheorem{defi}[propo]{Definition}
\newtheorem{conj}[propo]{Conjecture} 
\numberwithin{equation}{section}

\let\Diam\diamondsuit

\newcommand{\surj}{\to\kern-.1em\llap{$\to$}}

\newcommand{\jrus}{\leftarrow\kern-.1em\llap{$\leftarrow$}}
\newcommand{\strictsubset}{\hbox{$\subseteq\kern-.4em\llap{${}_/$}$}}

\newcommand{\norm}[1]{\lVert#1\lVert}
\newcommand{\plus}{\!+\!}

\DeclareMathOperator{\Exp}{Exp}

\begin{document}
\title{{Discrete period matrices and related topics}}
\author{Christian  \textsc{Mercat}}
\email{
\href{mailto:Christian.Mercat@entrelacs.net}{C.Mercat@ms.unimelb.edu.au}} 
\address{Department of Mathematics and Statistics\\
University of Melbourne\\ Parkville, Victoria 3010, Australia}
\begin{abstract}
  We continue our investigation of Discrete Riemann Surfaces with the
  discussion of the discrete analogs of period matrices, Riemann's bilinear
  relations, exponential of constant argument, series and electrical moves.
  We show that given a refining sequence of critical maps, the discrete
  period matrix converges to the continuous one.
\end{abstract}
\maketitle

\section{Introduction}
The notion of discrete Riemann surfaces was defined in~\cite{M,M01}. 
The interesting paper~\cite{CSMcC} initiated a renewed interest in the
subject.  In their paper, R.~Costa-Santos and B.~McCoy observed
numerically that certain pfaffians intervening in a dimer or critical
Ising model converge at the thermodynamic limit to a certain (power of
a) theta function at the origin.  They computed the period matrix
needed to define the theta function using discrete holomorphy method. 
The present paper aims at putting their work in a more general
theoretical framework.  Most of the results in this paper are a
straightforward application of the continuous theory~\cite{FK,Mum} together
with the results in~\cite{M,M01,M0206041}, to which we refer for details.  We
define the discrete period matrix, which is twice as large as in the
continuous case: the periods of a holomorphic form on the graph and on its
dual are in general different, but the continuous limit theorem, given a
refining sequence of critical maps, ensures that they converge to the same
value.  The main tool is the same as in the continuous case, the Riemann
bilinear relations.
We define the discrete exponential of a constant argument
on a critical map and explore its properties as well as its link with
series.  The tools for that purpose, which needs more investigations
in its own right, are the electrical moves.

\section{Discrete Riemann surfaces} \label{sec:Definitions}

We recall in this section basic definitions and results from~\cite{M01} where
the notion of discrete Riemann surfaces was defined.  We are interested in
discrete surfaces given by a cellular decomposition $\diamondsuit$ of
dimension two, where all faces are \emph{quadrilaterals} (a
quad-graph~\cite{BoS}).  It defines, up to homotopy and away from the
boundary, two dual cellular decompositions $\Gamma$ and $\Gamma^*$.
\begin{figure}[htbp]
\begin{center}\input{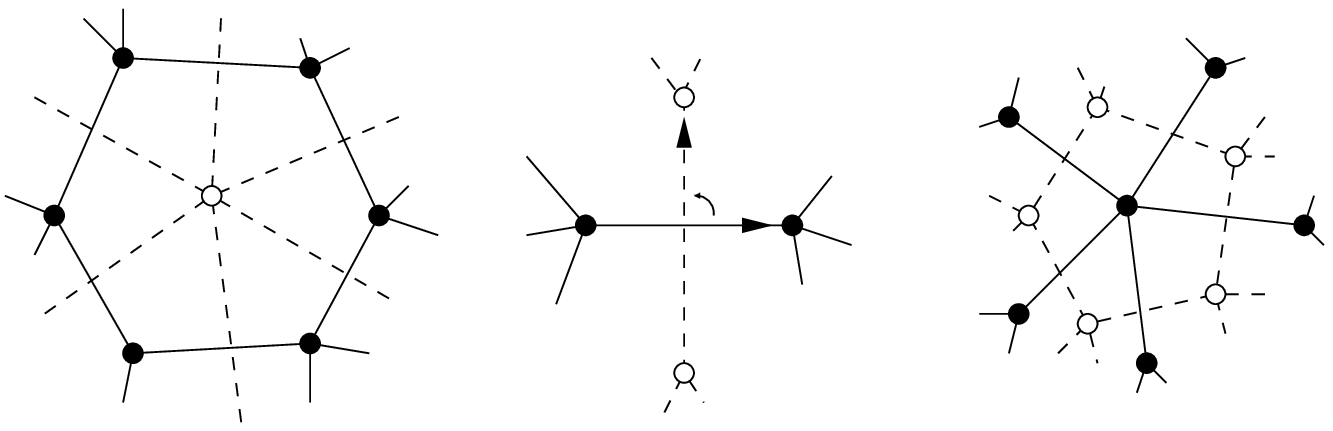}
\end{center}
\caption{Duality.}      \label{fig:duality}
\end{figure}
Edges $\Gamma^*_1$ are dual to edges $\Gamma_1$, faces $\Gamma^*_2$ are dual
to vertices $\Gamma_0$ and vice-versa.  Their union is denoted the
\emph{double} $\Lambda=\Gamma\sqcup \Gamma^*$. A \emph{discrete conformal
  structure} on $\Lambda$ is a real positive function $\rho$ on the
unoriented edges satisfying $\rho(e^*)=1/\rho(e)$. It defines a genuine
Riemann surface structure on the discrete surface: Choose a length $\delta$
and realize each quadrilateral by a lozenge whose diagonals have a length
ratio given by $\rho$.  Gluing them together provides a flat riemannian
metric with conic singularities at the vertices, hence a conformal
structure~\cite{Tro}.  It leads to a straightforward discrete version of the
\emph{Cauchy-Riemann equation}.  A function on the vertices is discrete
holomorphic iff for every quadrilateral $(x,y,x',y')\in\diamondsuit_2$,
\begin{equation}
  \label{eq:CR}
  f(y')-f(y)=i\,\rho(x,x')\left( f(x')-f(x)\right).
\end{equation}
\begin{figure}[htbp]
\begin{center}\input{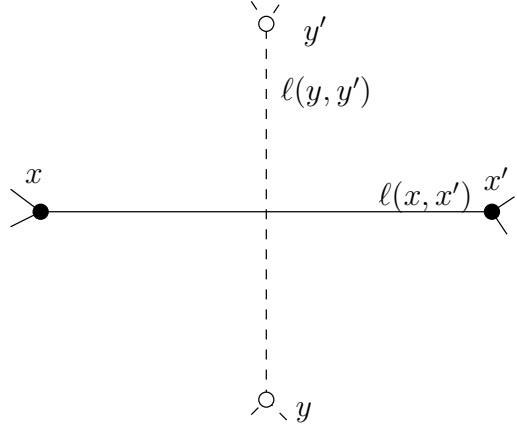}
\end{center}
\caption{The discrete Cauchy-Riemann equation.}         \label{fig:CR}
\end{figure}

We recall elements of de-Rham cohomology, doubled in our context: The complex
of \emph{chains} $C(\Lambda)=C_{0}(\Lambda)\oplus C_{1}(\Lambda)\oplus
C_{2}(\Lambda)$ is the vector space span by vertices, edges and faces.  It is
equipped with a \emph{boundary} operator $\partial:C_{k}(\Lambda)\to
C_{k-1}(\Lambda)$, null on vertices and fulfilling $\partial^{2}=0$.  The
kernel $\text{ker~}\partial=:Z_{\bullet}(\Lambda)$ of the boundary operator
are the closed chains or \emph{cycles}.  Its image are the \emph{exact}
chains.  It provides the dual spaces of forms, called \emph{cochains},
$C^{k}(\Lambda):=\text{Hom}(C_{k}(\Lambda),\mathbb{C})$ with a
\emph{coboundary} $d:C^k(\Lambda)\to C^{k+1}(\Lambda)$ defined by  Stokes
formula: $$\int\limits_{(x,x')} df:= f\left(\partial(x,x')\right)=f(x')-f(x),
\qquad \iint\limits_F d\alpha:=\oint\limits_{\partial F}\alpha.
$$
A \emph{cocycle} is a closed cochain and we note $\alpha\in 
Z^{k}(\Lambda)$.

Duality of complexes allows us to define a \emph{Hodge operator}
$*$ on forms by
\begin{eqnarray}
    *:C^{k}(\Lambda) & \to & C^{2-k}(\Lambda)
    \notag  \\
   \phantom{*:}C^{0}(\Lambda)\ni f & \mapsto & * f : \iint_{F}*f := f(F^{*}),
    \notag  \\
     \phantom{*:}C^{1}(\Lambda)\ni\alpha & \mapsto & * \alpha :
     \int_{e}*\alpha := -\rho(e^{*})\int_{e^{*}}\alpha,
    \label{eq:*Def}  \\
     \phantom{*:}C^{2}(\Lambda)\ni \omega & \mapsto & * \omega : 
     (*\omega)(x) := \iint_{x^{*}}\omega.
    \notag  
\end{eqnarray}

It fulfills $*^{2}=(-\text{Id})^{k}$ and defines $\Delta:=-d*d*-*d*d$, the
usual discrete \emph{Laplacian}. Its kernel are the harmonic forms. The
discrete holomorphic forms are special harmonic forms: a $1$-form
\begin{equation}
\alpha\in C^{1}(\Lambda) \text{~~is \emph{holomorphic} iff~~}
d\alpha=0 \text{~~and~~} *\alpha = -i\alpha, 
\label{eq:holoDef}
\end{equation}
that is to say if it is closed and of type $(1,0)$.  We will note
$\alpha\in\Omega^{1}(\Lambda)$.  A function
$f:\,\Lambda_{0}\to\mathbb{C}$ is \emph{holomorphic} iff $df$ is
holomorphic and we note $f\in\Omega^{0}(\Lambda)$.  

In the compact case, the Hodge theorem orthogonally decomposes forms into
exact, coexact and harmonic; harmonic forms are the closed and co-closed
ones; and harmonic $1$-form are the orthogonal sum of holomorphic and
anti-holomorphic ones. We studied discrete Hodge theory for higher
dimensional complexes in~\cite{M}.

We construct a wedge product on $\Diam$ such that $d_{\Diam}$ is a
derivation for this product $\wedge:C^k(\Diam)\times C^l(\Diam)\to
C^{k+l}(\Diam)$. It is defined by the following formulae, for $f,g\in
C^0(\Diam)$, $\alpha,\beta\in C^1(\Diam)$ and $\omega\in C^2(\Diam)$:
\begin{align*}
  (f\cdot g)(x):=&f(x)\cdot g(x)\qquad \mathrm{ ~for~} x\in\Diam_0,\\ 
  \int_{(x,y)}f\cdot\alpha:=& \frac{f(x)+f(y)}2
  \int{(x,y)}\alpha\qquad \mathrm{~for~} (x,y)\in\Diam_1,\\ 
  \iint\limits_{\hidewidth{(x_1,x_2,x_3,x_4)}\hidewidth}\alpha\wedge\beta
:=&\frac{1}{4}\sum_{k=1}^4
\int\limits_{{(x_{k-1},x_k)}}\alpha\;
\int\limits_{\hidewidth{(x_k,x_{k+1})}\hidewidth}\beta-
\int\limits_{{(x_{k+1},x_k)}}\alpha\;
\int\limits_{\hidewidth{(x_k,x_{k-1})}\hidewidth}\beta\\  
\iint\limits_{\hidewidth{(x_1,x_2,x_3,x_4)}\hidewidth}
  f\cdot\omega:=&\frac{\scriptstyle f(x_1)+f(x_2)+f(x_3)+f(x_4)}{4}
  \iint\limits_{\hidewidth{(x_1,x_2,x_3,x_4)}\hidewidth}\omega\\ 
  &\qquad \mathrm{~for~} (x_1,x_2,x_3,x_4)\in\Diam_2.
\end{align*}

\begin{figure}[htbp]
\begin{center}\input{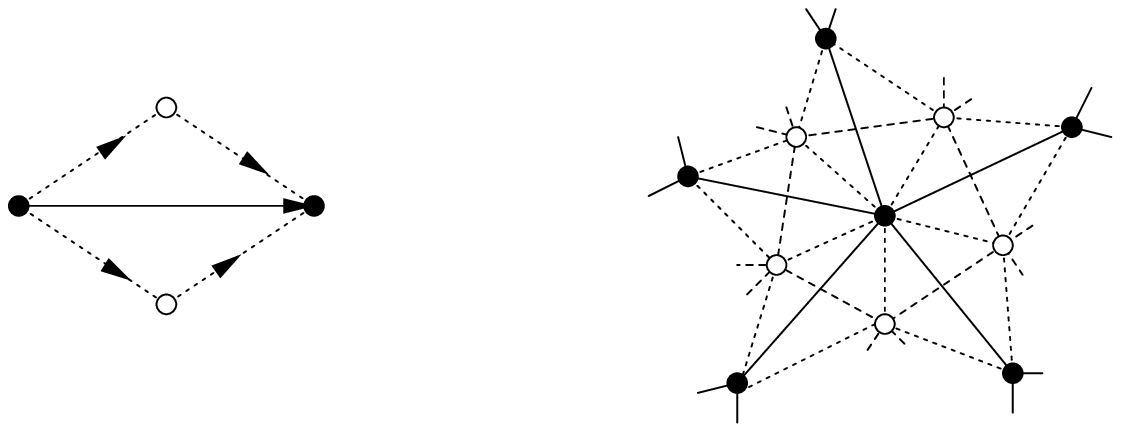}
\end{center}
\caption{Notations.}    \label{fig:avera}
\end{figure}

A form on $\Diam$ can be \emph{averaged} into a form on $\Lambda$:
This map $A$ from 
$C^\bullet(\Diam)$
to $C^\bullet(\Lambda)$ is the identity for functions and
defined by the following formulae for $1$ and $2$-forms:
\begin{align}
 \int\limits_{{(x,x')}}A(\alpha_\Diam):= \frac12 \left(
   \int\limits_{{(x,y)}}+\int\limits_{\hidewidth{(y,x')}\hidewidth}
   +\int\limits_{\hidewidth{(x,y')}\hidewidth}+\int\limits_{{(y',x')}}
 \right) \alpha_\Diam, \label{def:avera1}\\ 
 \iint\limits_{\hidewidth{x^*}\hidewidth}A(\omega_\Diam):=
\frac12\sum_{k=1}^d\;\iint\limits_{{(x_k,y_k,x,y_{k-1})}}\omega_\Diam,
\label{def:avera2}
\end{align}
where notations are made clear in Fig.~\ref{fig:avera}.  The map $A$ is
neither injective nor surjective in the non simply-connected case, so we can
neither define a Hodge star on $\Diam$ nor a wedge product on $\Lambda$.  Its
kernel is $\text{\rm Ker }(A)=\text{\rm Vect }(d_\Diam\varepsilon)$, where
$\varepsilon$ is the biconstant, $+1$ on $\Gamma$ and $-1$ on $\Gamma^*$.
But $d_\Lambda A=A\,d_\Diam$ so it carries cocycles on $\Diam$ to
cocycles on $\Lambda$ and its image are these cocycles of $\Lambda$ verifying
that their holonomies along cycles of $\Lambda$ only depend on their homology
on the combinatorial surface: Given a $1$-cocycle $\mu\in Z^{1}(\Lambda)$
with such a property, a corresponding $1$-cocycle $\nu\in Z^{1}(\Diam)$ is
built in the following way, choose an edge $(x_{0},y_{0})\in\Diam_{1}$; for
an edge $(x,y)\in\Diam_{1}$ with $x$ and $x_{0}$ on the same leaf of
$\Lambda$, choose two paths $\lambda_{x,x_{0}}$ and $\lambda_{y_{0},y}$ on
the double graph $\Lambda$, from $x$ to $x_{0}$ and $y_{0}$ to $y$
respectively, and define
    \begin{equation}
     \int_{(x,y)}\nu:=\int_{\lambda_{x,x_{0}}}\mu+\int_{\lambda_{y_{0},y}}\mu
        -\oint_{[\gamma]}\mu
        \label{eq:holonomyProp}
    \end{equation}
    where
    $[\gamma]=[\lambda_{x,x_{0}}+(x_{0},y_{0})+\lambda_{y_{0},y}+(y,x)]$ is
    the class of the full cycle in the homology of the surface. Changing the
    base points change $\mu$ by a multiple of $d_\diamondsuit\varepsilon$.

   It follows in the compact case that the dimensions of the harmonic forms
   on $\Diam$ (the kernel of $\Delta A$) modulo $d\varepsilon$, as well as
   the harmonic forms on $\Lambda$ with same holonomies on the graph and on
   its dual, are twice the genus of the surface, as expected.  Unfortunately,
   the space $\text{Im~} A=\mathcal{H}^\perp\oplus\text{Im }d$ is not stable
   by the Hodge star $*$.  We can nevertheless define holomorphic $1$-forms
   on $\Diam$ but their dimension can be much smaller than in the continuous,
   namely the genus of the surface.  Criticality provides conditions which
   ensure that the space $*\text{Im }A$ is ``close'' to $\text{Im }A$.

We 
construct an
{\em heterogeneous} wedge product for $1$-forms: with $\alpha,\beta\in
C^1(\Lambda)$, define $\alpha\wedge\beta\in C^1(\Diam)$ by
\begin{equation}
    \iint\limits_{\hidewidth(x,y,x',y')\hidewidth}\alpha\wedge\beta:=
    \int\limits_{\hidewidth(x,x')\hidewidth}\alpha
    \int\limits_{\hidewidth(y,y')\hidewidth}\beta+
    \int\limits_{\hidewidth(y,y')\hidewidth}\alpha
    \int\limits_{\hidewidth(x',x)\hidewidth}\beta.
    \label{eq:averaging}
\end{equation}

It verifies $A(\alpha_\Diam)\wedge A(\beta_\Diam)=\alpha_\Diam\wedge
\beta_\Diam$, the first wedge product being between $1$-forms on
$\Lambda$ and the second between forms on $\Diam$.  The usual scalar
product on compactly supported forms on $\Lambda$ reads
as expected:
 \begin{equation}
     (\alpha,\beta):=\sum_{e\in\Lambda_1}\rho(e)\left(
     \int_e\alpha\right) \left(
     \int_e\bar\beta\right)=\iint\limits_{\Diam_2}\alpha\wedge*\bar\beta
     \label{eq:scalarProd}
 \end{equation}

\section{Period matrix} \label{sec:PeriodMatrix}
We use the convention of Farkas and Kra~\cite{FK}, chapter III, to
which we refer for details.  Consider $(\Diam,\rho)$ a discrete
compact Riemann surface.

\subsection{Intersection number, on $\Lambda$ and on $\diamondsuit$} \label{sec:Intersection}
In~\cite{M}, for a given simple (real) cycle $C\in Z_{1}(\Lambda)$, we
constructed a harmonic $1$-form $\eta_{C}$ such that $\oint_{A}\eta_{C}$
counts the algebraic number of times $A$ contains an edge dual to an edge of
$C$.  It is the solution of a Neumann problem on the surface cut open along
$C$.  It follows from standard homology technique that $\eta_{C}$ depends
only on the homology class of $C$ (all the cycles which differ from $C$ by an
exact cycle $\partial A$) and can be extended linearly to all cycles; it
fulfills, for a closed form $\theta$,
\begin{equation}
  \label{eq:dualityEta}
  \oint_C \theta=\iint_\diamondsuit \eta_C\wedge\theta,
\end{equation}
and a basis of the homology provides a dual basis of harmonic forms on
$\Lambda$.  Beware that if the cycle $C\in Z_{1}(\Gamma)$ is purely on
$\Gamma$, then this form ${\eta_{C}}_{|_\Gamma}=0$ is null on $\Gamma$.
 
\begin{figure}[htbp]
\begin{center}\input{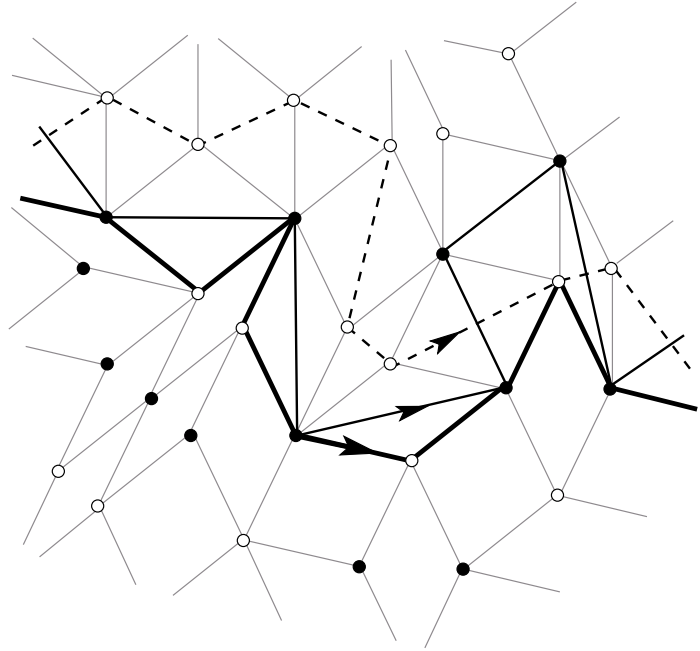}
\end{center}
\caption{A path $C$ on $\diamondsuit$ defines a pair of paths $C_{\Gamma}$ and  $C_{\Gamma^*}$ on its left.}
\label{fig:diamGamGams}
\end{figure}

The \emph{intersection number} between two cycles $A,B\in 
Z_{1}(\Lambda)$ is defined as
\begin{equation}
    A\cdot B:=\iint_{\Diam}\eta_{A}\wedge\eta_{B}.
    \label{eq:interDef}
\end{equation}
It is obviously linear and antisymmetric, it is an integer number for integer
cycles. Let's stress again that the intersection of a cycle on $\Gamma$ with
another cycle on $\Gamma$ is always null.  A cycle $C\in Z_{1}(\Diam)$
defines a pair of cycles on each graph $C_{\Gamma}\in Z_{1}(\Gamma)$,
$C_{\Gamma^{*}}\in Z_{1}(\Gamma^{*})$ which are homologous to $C$ on the
surface, composed of portions of the boundary of the faces on $\Lambda$ dual
to the vertices of $C$.  They are uniquely defined if we require that they
lie ``to the left'' of $C$ as shown in Fig.\ref{fig:diamGamGams}.
By the procedure~\eqref{eq:holonomyProp} applied to
$\eta_{C_{\Gamma}}+\eta_{C_{\Gamma^{*}}}$, we construct a $1$-cocycle
$\eta_{C}\in Z^{1}(\Diam)$ unique up to $d\varepsilon$, and since
$\forall\theta,\; d\varepsilon\wedge\theta=0$, Eq.~\eqref{eq:interDef}
defines an intersection number on $Z_{1}(\Diam)$.  Unlike the intersection
number on $\Lambda$, this one has all the usual expected properties. In
particular Eq.~\eqref{eq:interDef} holds for $A,B\in Z_{1}(\diamondsuit)$.

\subsection{Canonical dissection, fundamental polygon} \label{sec:CanonicalDiss}
The complex $\Diam$ being connected, consider a maximal tree
$T\subset\Diam_{1}$, that is to say $T$ is a $\mathbf{Z}_2$-homologically
trivial chain and every edge added to $T$ forms a cycle.  A \emph{canonical
  dissection} or cut-system $\aleph$ of the genus $g$ discrete Riemann
surface $\Diam$ is given by a set of oriented edges $(e_{k})_{1\leq k\leq
  2g}$ such that the cycles $\aleph\subset(T\cup e_{k})$ form a basis of the
homology group $H_{1}(\Diam)$ verifying, for $1\leq k,\ell\leq g$
\begin{equation}
    \aleph_{k}\cdot\aleph_{\ell}=0,   \quad
    \aleph_{k+g}\cdot\aleph_{\ell+g}=0,   \quad
    \aleph_{k}\cdot\aleph_{\ell+g}=\delta_{k,\ell}.
 \label{eq:alephDiam} 
\end{equation}
They actually form a basis of the fundamental group
$\pi_1(\diamondsuit)$ and the defining relation among them is (noted
multiplicatively)
\begin{equation}
  \label{eq:pi1}
  \prod_{k=1}^g\aleph_{k}\aleph_{k+g}\aleph_{k}^{-1}\aleph_{k+g}^{-1}=1.
\end{equation}
The construction of such a basis is standard and we won't repeat the
procedure. What is less standard is the interpretation of Eq.~\eqref{eq:pi1}
in terms of the boundary of a fundamental domain, discretization introduces
some subtleties.  

\begin{figure}[htbp]
\begin{center}\input{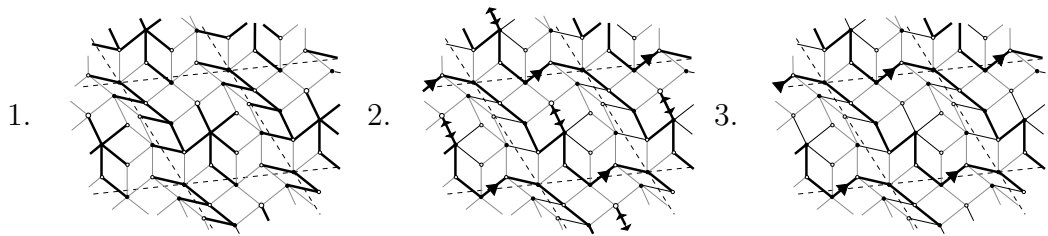}
\end{center}
\caption{1. A maximal rooted tree in a quadrilateral decomposition of the
  torus. 2. An additional edge defines a rooted cycle $\aleph_1$, pruned of
  its dangling trees. 3. Its un-rooted version, the simple loop $\aleph_1^-$.}
\label{fig:prune}
\end{figure}

Considering $T\cup e_{k}$ as a rooted graph, we can prune it of all its
pending branches, leaving a simple closed loop $\aleph_k^-$, attached to the
origin $O$ by a simple path $\lambda_k$ (see Fig.~\ref{fig:prune}), yielding
the cycle $\aleph_k$. These three cycles are deformation retract of one
another, $\aleph_k^-\subset \aleph_k\subset T\cup e_{k}$ hence are equal in
homology.

In the continuous case, a basis of the homology can be realized by $2g$
simple arcs, transverse to one another and meeting only at the base point. It
defines an isometric model of the surface as a fundamental domain
homeomorphic to a disc and bordered by $4g$ arcs to identify pairwise.  In
the discrete case, by definition, the set $\diamondsuit\setminus\aleph$ of
the cellular complex minus the edges taking part into the cycles basis is
homeomorphic to a disc hence the surface is realized as a polygonal
\emph{fundamental domain} $\mathcal{M}$ whose boundary edges are identified
pairwise.

But it is sometimes impossible to choose a basis of the homology
verifying~\eqref{eq:alephDiam} by simple discrete cycles which are transverse
to one another. If the path $\lambda_k$ is not empty, the cycle $\aleph_k$ is
not even simple. Moreover, some edges may belong to several cycles. In this
case, the edges on the boundary of this fundamental polygon can not be
assigned a unique element of the basis or its inverse, and therefore can not
be grouped into only $4g$ continuous paths to identify pairwise but more than
$4g$.

\begin{figure}[htbp]
\begin{center}\input{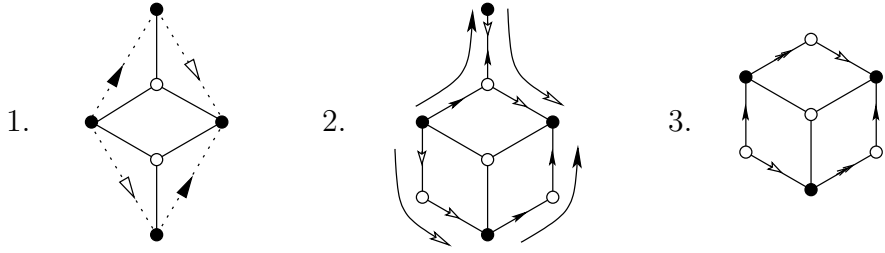}
\end{center}
\caption{Three different fundamental polygons of a decomposition of the torus
  ($g=1$) by three quadrilaterals: 1. The standard fundamental domain where
  the $4g$ paths are not adapted to $\diamondsuit$. 2. $\mathcal{M}_+$ is
  composed of edges of $\diamondsuit$ composing $4g$ arcs (which may have
  portions in common) to identify pairwise, each edge corresponds to an
  element of the basis $\aleph$ or its inverse, except for edges of
  ``dangling trees'' which are associated with two such elements.  3.
  $\mathcal{M}$ is composed of edges of $\diamondsuit$ composing more than
  $4g$ arcs to identify pairwise, there is no correspondence with a basis of
  cycles.}
\label{fig:FundPoly}
\end{figure}

In fact, the information contained into the basis $\aleph$ is more than
simply this polygon, the set of edges composing the concatenated cycle
\begin{equation}
  \label{eq:FundPoly1}
(\aleph_1,\aleph_{g+1},\aleph_1^{-1},\aleph_{g+1}^{-1},\aleph_2,\ldots,
\aleph_{g}^{-1},\aleph_{2g}^{-1})
\end{equation}
encodes a cellular complex $\mathcal{M}_+$ which is \emph{not} a
combinatorial surface and consists of the fundamental polygon $\mathcal{M}$
plus some \emph{dangling trees}, corresponding to the edges which belong to
more than one cycle or participate more than once in a cycle (the paths
$\lambda_k$), as exemplified in Fig.\ref{fig:FundPoly}. By construction, the
edge $e_k$ belongs to the cycle $\aleph_k$ only, hence these trees are in
fact without branches, simple paths whose only leaf is the base point $O$. To
retrieve the surface, the edges of this structure $\mathcal{M}_+$ are
identified group-wise, an edge participating $k$ times in cycles will have
$[k/2]+2$ representatives to identify together, two on the fundamental
polygon and the rest as edges of dangling trees.

Eliminating repetition, that is to say looking at~\eqref{eq:FundPoly1} not as
a \emph{sequence} of edges but as a simplified \emph{cycle} (or a simplified
word in edges), thins $\mathcal{M}_+$ into $\mathcal{M}$, pruning away the
dangling paths.  The fundamental polygon boundary loses its structure as $4g$
arcs to be identified pairwise, in general a basis cycle will be disconnected
around the fundamental domain and a given edge can not be assigned to a
particular cycle.  This peculiarity gives a more complex yet well defined
meaning to the contour integral formula for a $1$-form $\theta$ defined on
the boundary edges of $\mathcal{M}_+$,
\begin{equation}
    \oint_{\partial\mathcal{M}}\theta=\sum_{k=1}^{2g}
    \oint_{\aleph_{k}}\theta+\oint_{\aleph_{k}^{-1}}\theta.
    \label{eq:contourInt}
\end{equation}
This basis gives rise to cycles $\aleph^{\Gamma}$ and $\aleph^{\Gamma^{*}}$
whose homology classes form a basis of the group for each respective graph,
that we compose into $\aleph^{\Lambda}$ defined by
\begin{align}
    \aleph^{\Lambda}_{k} & =  \aleph^{\Gamma}_{k},
      &
    \aleph^{\Lambda}_{k+g} & =  \aleph^{\Gamma^{*}}_{k},
    \label{eq:alephLambda}  \\
    \aleph^{\Lambda}_{k+2g} & =  \aleph^{\Gamma^{*}}_{k+g},
     &
    \aleph^{\Lambda}_{k+3g} & =  \aleph^{\Gamma}_{k+g},
    \notag
\end{align}
for $ 1\leq k\leq g $ so that while the intersection numbers matrix on
$\diamondsuit$ is given by the $2g\times 2g$ matrix
\begin{equation}
    (\aleph^{}_{k}\cdot\aleph^{}_{\ell})_{k,\ell}=
    \begin{pmatrix}
        0 & I  \\
        -I & 0  
    \end{pmatrix},
    \label{eq:alephIntersecDiam}
\end{equation} 
the intersection numbers matrix on $\Lambda$ is the $4g\times 4g$ matrix with
the same structure
\begin{equation}
    (\aleph^{\Lambda}_{k}\cdot\aleph^{\Lambda}_{\ell})_{k,\ell}=
    \begin{matrix}
      \begin{matrix}
       \phantom{-}\Gamma^{ } &\phantom{-}\Gamma^* &
\!\Gamma^* & \!\Gamma^{ }
      \end{matrix}\\
      \begin{pmatrix}
0&0&
      I & 0  \\
0&0&
      0 & I
\\
     -I & 0   &0&0\\
      0 &  -I   &0&0
  \end{pmatrix}&
  \begin{matrix}
    \Gamma^{\phantom{*}}\\
    \Gamma^*\\
    \Gamma^*\\
    \Gamma^{\phantom{*}}
  \end{matrix}
  \end{matrix}
.
    \label{eq:alephIntersec}
\end{equation}

\subsection{Basis of harmonic forms, basis of holomorphic forms} \label{sec:Basis}
We define $\alpha^{\Lambda}$, the basis of real harmonic $1$-forms, dual to
the homology basis $\aleph^{\Lambda}$,
\begin{eqnarray}
\alpha^{\Lambda}_{k}&:=&\eta_{\aleph^{\Lambda}_{k+2g}} \qquad 
\text{ ~and}\notag\\
\alpha^{\Lambda}_{k+2g}&:=&-\eta_{\aleph^{\Lambda}_{k}} \qquad 
\text{ ~for~ } 1\leq k\leq 2g
\label{eq:dualBasisDef}
\end{eqnarray}
which verify
\begin{eqnarray}
\oint_{\aleph^{\Lambda}_{k}}\alpha_{\ell}&=&\delta_{k,\ell},\notag\\
\oint_{\aleph^{\Lambda}_{k+2g}}\alpha_{\ell+2g}&=&\delta_{k,\ell},
\label{eq:dualBasis}
\end{eqnarray}
and dually, the intersection matrix elements are given by
\begin{equation}
    \aleph^{\Lambda}_{k}\cdot\aleph^{\Lambda}_{\ell}= 
    \iint_{\Diam}\alpha^{\Lambda}_{k}\wedge\alpha^{\Lambda}_{\ell}=
    (\alpha^{\Lambda}_{k},-*\alpha^{\Lambda}_{\ell}).  
    \label{eq:alephIntersecAlpha}
\end{equation}
On $\Diam$, the elements $\alpha^{\Diam}_{k}:=\eta_{\aleph_{k+g}}$ and
$\alpha^{\Diam}_{k+g}:=-\eta_{\aleph_{k}}$ for $1\leq k\leq g$, defined up to
$d\varepsilon$, verify
$A(\alpha^{\Diam}_{k})=\alpha^{\Lambda}_{k}+\alpha^{\Lambda}_{k+g}$,
$A(\alpha^{\Diam}_{k+g})=\alpha^{\Lambda}_{k+2g}+\alpha^{\Lambda}_{k+3g}$ and
form a basis of the cohomology on $\Diam$ dual to $\aleph$ as well,
\begin{eqnarray}
\alpha^{\diamondsuit}_{k}&:=&\eta_{\aleph^{\diamondsuit}_{k+g}} \qquad 
\text{ ~and}\notag\\
\alpha^{\diamondsuit}_{k+g}&:=&-\eta_{\aleph^{\diamondsuit}_{k}} \qquad 
\text{ ~for~ } 1\leq k\leq g,
\label{eq:dualBasisDefDiam}
\end{eqnarray}
they fulfill the first identity in~Eq.\eqref{eq:alephIntersecAlpha} but the
second is meaningless in general since $*$ can not be defined on
$\diamondsuit$. We will drop the mention $\Lambda$ when no confusion is
possible.

\begin{propo} \label{prop:positive}
  The matrix of inner products on $\Lambda$,
 \begin{equation}
     (\alpha_{k},\alpha_{\ell})_{k,\ell}
     =  \iint_{\Diam}\alpha_{k}\wedge*\bar\alpha_{\ell}
     = \begin{cases}
        +\oint_{\aleph_{k+2g}} *\alpha_\ell,& 1\leq k\leq 2g,\\
        -\oint_{\aleph_{k-2g}} *\alpha_\ell,& 2g< k\leq 4g.
       \end{cases}
   =:
     \begin{pmatrix}
       A & D  \\
       B & C
     \end{pmatrix}
     \label{eq:GammaMatrix}
 \end{equation}  
  is a real symmetric positive definite matrix.
\end{propo}
\begin{demo}{\ref{prop:positive}}
  It is real because the forms are real, and symmetric because the scalar
  product~\eqref{eq:scalarProd} is skew symmetric. Definition
  Eq.~\eqref{eq:dualBasisDef} and Eq.~\eqref{eq:dualityEta} lead to the
  integral formulae.  Positivity follows from the bilinear
  relation~Eq.~\eqref{eq:iintTheta}: for
  $\theta=\sum_{k=1}^{4g}\xi_k\,\alpha_k$, with
  $\xi_k\in\mathbf{C},\;\sum_{k=1}^{4g}|\xi_k|^2>0$,
\begin{eqnarray}
  \norm{\theta}^2&=&\sum_{j=1}^{2g}\left[
\int_{\aleph_j}\theta\int_{\aleph_{2g+j}}*\bar\theta-\int_{\aleph_{2g+j}}\theta\int_{\aleph_{2j}}*\bar\theta\right]\notag\\
&=&\sum_{k,\ell=1}^{4g}\xi_k\,\bar\xi_\ell\sum_{j=1}^{2g}\left[
\int_{\aleph_j}\alpha_k\int_{\aleph_{2g+j}}*\alpha_\ell
-\int_{\aleph_{2g+j}}\alpha_k\int_{\aleph_{2j}}*\alpha_\ell\right]\notag\\
&=&\sum_{k,\ell=1}^{4g}\xi_k\,\bar\xi_\ell\,(\alpha_{k},\alpha_{\ell})>0.
  \label{eq:positive}
\end{eqnarray}
\end{demo}

The form $\alpha_k$ is supported by only one of the two graphs $\Gamma$ or
$\Gamma^{*}$, the form $* \alpha_k$ is supported by the other one, and the
wedge product $\theta_\Gamma\wedge\theta'_\Gamma=0$ is null for two $1$-forms
supported by the same graph. Therefore the matrices $A$ and $C$ are $g\times
g$-block diagonal and $B$ is anti-diagonal. 
\begin{equation}
  \label{eq:blockDiag}
  A=
  \begin{pmatrix}
    A_\Gamma&0\\ 0&A_{\Gamma^*}
  \end{pmatrix},\qquad
  B=
  \begin{pmatrix}
   0&B_{\Gamma^*,\Gamma}\\ B_{\Gamma,\Gamma^*}&0
  \end{pmatrix},\qquad
  C=
  \begin{pmatrix}
   C_{\Gamma^*}&0\\ 0&C_{\Gamma}
  \end{pmatrix}.
\end{equation}
The matrices of intersection numbers~\eqref{eq:alephIntersec} and of inner
products differ only by the Hodge star $*$. Because $*$ preserves harmonic
forms and the inner product, we get its matrix representation in the basis
$\alpha$,
\begin{equation}
  \label{eq:*Mat}
  *=
  \begin{pmatrix}
    -D&A\\
    -C&B
  \end{pmatrix}
\end{equation}
and because $*^2=-1$,
\begin{eqnarray}
B^2-C\cdot A+I&=&0\\
A\cdot B&=&{}^tB\cdot A\\
C\cdot {}^tB&=&B\cdot C.
\end{eqnarray}

On $\diamondsuit$, while the Hodge star $*$ can not be defined, we can
obviously consider the following positive scalar product on the classes of
closed forms modulo $d\varepsilon$, to which the set
$(\alpha^\diamondsuit_k)$ belong:
\begin{eqnarray*}
\notag
  (\alpha^\diamondsuit,\beta^\diamondsuit)&:=&
\left(A(\alpha^\diamondsuit),A(\beta^\diamondsuit)\right)
\\
 &=&
\sum_{\substack{(x,y,x',y')\in\diamondsuit_2\\
 \rho=\rho(x,x'),\,\rho^*=\rho(y,y')}}
\begin{pmatrix}
{\hbox to 0pt{\kern -1.3em t}} \int_{(x\phantom{'},y\phantom{'})}\alpha\\  
\int_{(y\phantom{'},x')}\alpha\\ 
\int_{(x',y')}\alpha\\ \int_{(y',x\phantom{'})}\alpha
\end{pmatrix}
\cdot
\begin{pmatrix}
\scriptscriptstyle +\rho+\rho^*&\scriptscriptstyle+\rho-\rho^*&
\scriptscriptstyle -\rho-\rho^*&\scriptscriptstyle-\rho+\rho^*\\
\scriptscriptstyle +\rho-\rho^*&\scriptscriptstyle+\rho+\rho^*&
\scriptscriptstyle -\rho+\rho^*&\scriptscriptstyle-\rho-\rho^*\\
\scriptscriptstyle -\rho-\rho^*&\scriptscriptstyle-\rho+\rho^*&
\scriptscriptstyle +\rho+\rho^*&\scriptscriptstyle+\rho-\rho^*\\
\scriptscriptstyle -\rho+\rho^*&\scriptscriptstyle-\rho-\rho^*&
\scriptscriptstyle +\rho-\rho^*&\scriptscriptstyle+\rho+\rho^*
\end{pmatrix}
\cdot
\begin{pmatrix}
\int_{(x\phantom{'},y\phantom{'})}\bar\beta\\  
\int_{(y\phantom{'},x')}\bar\beta\\ 
\int_{(x',y')}\bar\beta\\ \int_{(y',x\phantom{'})}\bar\beta
\end{pmatrix}.
\notag  \label{eq:scalarProdDiam} 
\end{eqnarray*}
and it yields
 \begin{equation}
     (\alpha^\diamondsuit_{k},\alpha^\diamondsuit_{\ell})_{k,\ell}
   =
     \begin{pmatrix}
       A_\Gamma+A_{\Gamma^*} & 
{}^tB_{\Gamma\Gamma^*}+{}^tB_{\Gamma^*\Gamma}  \\
      B_{\Gamma\Gamma^*}+B_{\Gamma^*\Gamma} & C_\Gamma+C_{\Gamma^*}
     \end{pmatrix},
     \label{eq:GammaMatrixDiam}
 \end{equation}  
 which, in general, can not be understood as the periods of a set of forms on
 $\diamondsuit$ along the basis $\aleph$.

Let's decompose the space of harmonic forms into two orthogonal supplements,
   \begin{equation}
     \label{eq:parallelPerp}
     \mathcal{H}^1(\Lambda)=
\mathcal{H}^1_\parallel\oplus^\perp \mathcal{H}^1_\perp
   \end{equation}
where the first vector space are the harmonic forms whose holonomies on one
graph are equal to their holonomies on the dual, that is to say
\begin{equation}
  \label{eq:parallel}
  \mathcal{H}^1_\parallel:=\text{Vect }(\alpha_k+\alpha_{k+g}, \;
1\leq k\leq g \text{ ~or~ } 2g< k\leq 3g).
\end{equation}
Definition \eqref{eq:dualBasisDef} and Eq.~\eqref{eq:dualityEta} imply that
\begin{equation}
  \label{eq:perp}
  \mathcal{H}^1_\perp=\text{Vect }(*\alpha_k-*\alpha_{k+g}, \;
1\leq k\leq g \text{ ~or~ } 2g< k\leq 3g).
\end{equation}
These elements in the basis $(\alpha_k+\alpha_{k+g}, \; ;\;
\alpha_k-\alpha_{k+g})$ for $1\leq k\leq g$ and $2g< k\leq 3g$,
are represented by the following invertible matrix:
\begin{equation}
  \label{eq:matrix+-}
  \begin{pmatrix}
    I&0&{}^tB_{\Gamma\Gamma^*}-{}^tB_{\Gamma^*\Gamma}&A_\Gamma-A_{\Gamma^*}\\
    0&I&C_\Gamma-C_{\Gamma^*}&B_{\Gamma\Gamma^*}-B_{\Gamma^*\Gamma}\\
0&0&{}^tB_{\Gamma\Gamma^*}+{}^tB_{\Gamma^*\Gamma}&A_\Gamma+A_{\Gamma^*}\\
0&0&C_\Gamma+C_{\Gamma^*}&B_{\Gamma\Gamma^*}+B_{\Gamma^*\Gamma}
  \end{pmatrix}.
\end{equation}
It implies in particular that the lower right $g\times g$ block is
invertible, therefore so is Eq.~\eqref{eq:GammaMatrixDiam}.

\subsection{Period matrix} \label{sec:periodMatrix}
 \begin{propo} \label{prop:Period}
   The matrix $\Pi=C^{-1}\cdot(i-B)$ is the \emph{period matrix} of the basis
   of holomorphic forms
 \begin{equation}
     \zeta_{k}:=(i-*)\sum_{\ell=1}^{2g}C^{-1}_{k,\ell}\;\alpha_{\ell+2g}
     \label{eq:zetaDef}
 \end{equation}  
 in the canonical dissection $\aleph$, that is to say
 \begin{equation}
     \oint_{\aleph_{k}}\zeta_{\ell}=
     \begin{cases}
         \delta_{k,\ell} & \text{~ for ~} 1\leq 
     k\leq 2g, \\
     \Pi_{k-2g,\ell} & \text{~ for ~} 2g< 
         k\leq 4g,
     \end{cases}
     \label{eq:zetaDual}
 \end{equation}
 and $\Pi$ is symmetric, with a positive definite imaginary part.
 \end{propo}
 The proof is essentially the same as in the continuous case~\cite{FK} and we
 include it for completeness.

 \begin{demo}{\ref{prop:Period}}
   Let $\omega_j:=\alpha_j+i*\alpha_j$ for $1\leq j\leq 4g$. These
   holomorphic forms fulfill
   \begin{eqnarray}
     \label{eq:omegaj}
     P_{k,j}:=\frac{1}{2}(\omega_k,\omega_j)&=&
(\alpha_k,\alpha_j)+i\,(\alpha_k,-*\,\alpha_j)\\
&=&
     \begin{cases}
        -i\,\int_{\aleph_{j+2g}} \omega_k,& 1\leq j\leq 2g,\\
         i\,\int_{\aleph_{j-2g}} \omega_k,& 2g< j\leq 4g.
     \end{cases}
   \end{eqnarray}
P is the period matrix of the forms $(\omega)$ in the homology basis $\aleph$.
The first $2g$ forms $(\omega_j)_{1\leq j\leq 2g}$ are a basis of holomorphic
forms. It has the right dimension and they are linearly independent: 
\begin{eqnarray}
  \sum_{j=1}^{2g}(\lambda_j+i\mu_j)(\alpha_j+i*\,\alpha_j)&=&
 \sum_{j=1}^{2g}\left((\lambda_j+\sum_{k=1}^{2g}\mu_k\,
 B_{j,k})\,\alpha_j
+\sum_{k=1}^{2g}\mu_k\, C_{j,k}\,\alpha_{2g+j}\right)\notag\\
&&\; +i\,\sum_{j=1}^{2g}\left((\mu_j+\sum_{k=1}^{2g}\lambda_k\,
 B_{j,k})\,\alpha_j+\sum_{k=1}^{2g}\lambda_k\, C_{j,k}\,\alpha_{2g+j}\right)
  \label{eq:linomega}
\end{eqnarray}
is null, for $\lambda,\mu\in\mathbf{R}$ only when $\lambda=\mu=0$ because $C$
is positive definite. Similarly for the last $2g$ forms. The change of basis
$i\,C^{-1}$ on them provides the basis of holomorphic forms $(\zeta)$. The
last $2g$ rows of $P$ is the $2g\times 4g$ matrix $(B-i\,I,C)$ hence the
periods of $(\zeta)$ in $\aleph$ are given by $(I,\Pi)$.
\end{demo}

The first identity in~Eq.\eqref{eq:zetaDual} uniquely defines the basis
$\zeta$ and a holomorphic $1$-form is completely determined by whether its
periods on the first $2g$ cycles of $\aleph$, or their real parts on the
whole set.

Notice that because $C$ is $g\times g$ block diagonal and $B$ is
anti-diagonal, $\Pi$ is decomposed into four $g\times g$ blocks, the two
diagonal matrices form $i\, C^{-1}$ and are pure imaginary, the other two
form $-C^{-1}\cdot B$ and are real.
 \begin{equation}
   \label{eq:PiBlocks}
   \Pi=
   \begin{pmatrix}
     \Pi_{i*}&\Pi_{r}\\
\Pi_{r*}&\Pi_{i}
   \end{pmatrix}
=
   \begin{pmatrix}
     i\,C_{\Gamma^*}^{-1}&-C_{\Gamma^*}^{-1}\cdot B_{\Gamma^*,\Gamma}\\
-C_{\Gamma}^{-1}\cdot B_{\Gamma,\Gamma^*}&i\,C_{\Gamma}^{-1}
   \end{pmatrix}.
 \end{equation}
 
 Therefore the holomorphic forms $\zeta_k$ are real on one graph and pure
 imaginary on its dual,
 \begin{eqnarray}
   \label{eq:zetaRI}
   1\leq k\leq g&\Rightarrow& \zeta_k\in
   C_{\mathbf{R}}^1(\Gamma)\oplus i\,C_{\mathbf{R}}^1(\Gamma^*)
\\
   g< k\leq 2g&\Rightarrow& \zeta_k\in
   C_{\mathbf{R}}^1(\Gamma^*)\oplus i\,C_{\mathbf{R}}^1(\Gamma).
\notag
 \end{eqnarray}
 
 We will call
 \begin{equation}
   \label{eq:PiGamma}
   \Pi_\Gamma=\Pi_r+\Pi_{i*}
 \end{equation}
 \emph{the period matrix on the graph $\Gamma$} the sum of the real periods
 of $\zeta_{k}$, $1\leq k\leq g$, on $\Gamma$, with the associated pure
 imaginary periods on the dual $\Gamma^{*}$, and similarly for $\zeta_{k}$,
 $g< k\leq 2g$, the period matrix on $\Gamma^{*}$.
 
 It is natural to ask how close $\Pi_{\Gamma}$ and $\Pi_{\Gamma^*}$ are from
 one another, and whether their mean can be given an
 interpretation. Criticality~\cite{M,M01} redefined in
 Sec.~\ref{sec:criticality}, answers partially the issue:

 \begin{theo}\label{th:critic}
   In the genus one critical case, the period matrices $\Pi_{\Gamma}$ and
   $\Pi_{\Gamma^*}$ are equal to the period matrix $\Pi_\Sigma$ of the
   underlying surface $\Sigma$. For higher genus, given a refining sequence
   $(\diamondsuit^k,\rho_k)$ of critical maps of $\Sigma$, the discrete
   period matrices $\Pi_{\Gamma^k}$ and $\Pi_{\Gamma^{*k}}$ converge to the
   period matrix $\Pi_\Sigma$.
 \end{theo}
 \begin{demo}{\ref{th:critic}}
   The genus one case is postponed to Sec.~\ref{sec:GenusOne}. The continuous
   limit comes from techniques in~\cite{M,M01}, to be developed
   in~\cite{M0206041} which prove that, given a refining sequence of critical
   maps, any holomorphic function can be approximated by a sequence of
   discrete holomorphic functions. Taking the real parts, this implies as
   well that any harmonic function can be approximated by discrete harmonic
   functions. In particular, the discrete solutions $f_k$ to a Dirichlet or
   Neumann problem on a simply connected set converge to the continuous
   solution $f$ because the latter can be approximated by discrete harmonic
   functions $g_k$ and the difference $f_k-g_k$ being harmonic and small on
   the boundary, converge to zero. In particular, each form in the basis
   $(\alpha^\diamondsuit_\ell)$, provides a solution to the Neumann problem
   Eq.~\eqref{eq:dualBasisDefDiam} and a similar procedure, detailed
   afterwards, define a converging sequence of forms
   $\zeta^\diamondsuit_\ell$, yielding the result.
 \end{demo}
 
 We can try to replicate the work done on $\Lambda$ on the graph
 $\diamondsuit$.  A problem is that $A_{\Gamma}+A_{\Gamma}$ and
 $C_{\Gamma}+C_{\Gamma}$ need not be positive definite. Moreover, the Hodge
 star $*$ doesn't preserve the space $(A(\alpha^{\Diam}_k))$ of harmonic
 forms with equal holonomies on the graph and on its dual, so we can not
 define the analogue of $\alpha+i\,*\alpha$ on $\diamondsuit$. We first
 investigate what happens when we can partially define these analogues:
 
 Assume that for $2g<k\leq 3g$, the holonomies of $*\alpha_k$ on $\Gamma$ are
 equal to the holonomies of $*\alpha_{k+g}$ on $\Gamma^*$, that is to say
 $C_\Gamma=C_{\Gamma^*}=:\frac{1}{2}C_\diamondsuit$ and
 $D_{\Gamma\Gamma^*}=D_{\Gamma^*\Gamma}=:\frac{1}{2}D_\diamondsuit$. It
 implies that the transposes fulfill
 $B_{\Gamma\Gamma^*}=B_{\Gamma^*\Gamma}=:\frac{1}{2}B_\diamondsuit$ as well.
 We can then define $\beta^{\Diam}_{k-g}\in Z^1(\diamondsuit)$ such that
 $A(\beta^{\Diam}_{k-g})=*\alpha_{k+g}$, uniquely up to $d\varepsilon$. The
 last $g$ columns ${}^t(B_\diamondsuit,C_\diamondsuit)$ of the matrix of
 scalar product Eq.~\eqref{eq:GammaMatrixDiam} are related to their periods
 in the homology basis $\aleph^{\Diam}$ in a way similar to
 Eq.~\eqref{eq:GammaMatrix}. By the same reasoning as before, the forms
\begin{equation}
  \label{eq:zetaDiam}
  \zeta^\diamondsuit_k=\sum_{\ell=1}^g{C_\diamondsuit^{-1}}_{k,\ell}
\left(\alpha^{\Diam}_{\ell+g}-i\beta^{\Diam}_{\ell+g}\right), \;\;
1\leq k\leq g
\end{equation}
verify $A(\zeta^\diamondsuit_k)=\frac{\zeta_k+\zeta_{k+g}}{2}$ and have
periods on $\aleph^{\Diam}$ given by the identity for the first $g$ cycles
and the following $g\times g$ matrix, mean of the period matrices on the
graph and on its dual:
\begin{equation}
\Pi^{\Diam}=C_\diamondsuit^{-1}(i-B_\diamondsuit)=
\frac{\Pi_{\Gamma}+\Pi_{\Gamma^*}}{2}.
     \label{eq:PeriodMatrixDiamLambda}
 \end{equation}
 
 The same reasoning applies when the periods of the forms $*\alpha_k$ on the
 graph and on its dual are not equal but close to one another. In the context
 of refining sequences, we said that the basis $(\alpha^\diamondsuit_\ell)$,
 converges to the continuous basis of harmonic forms defined by the same
 Neumann problem Eq.~\eqref{eq:dualBasisDefDiam}. Therefore
 \begin{equation}
   \label{eq:CGammaCGammaS}
   C_\Gamma-C_{\Gamma^*}=o(1), \qquad
 B_{\Gamma\Gamma^*}-B_{\Gamma^*\Gamma}=o(1).
 \end{equation}
 A harmonic form $ \nu_{k+g}=o(1) $ on $\Gamma^*$ can be added to $
 *\alpha_{k+g} $ such that there exists $ \beta^{\Diam}_{k-g}\in
 Z^1(\diamondsuit) $ with $ A(\beta^{\Diam}_{k-g})=*\alpha_{k+g}+\nu_{k+g}$,
 yielding forms $ \zeta^\diamondsuit_k$, verifying $
 A(\zeta^\diamondsuit_k)=\frac{1}{2}(\zeta_k+\zeta_{k+g})+o(1) $ and whose
 period matrix is $ \Pi^{\Diam}+o(1)$. Since the periods of $\alpha_k$
 converge to the same periods as its continuous limit, this period matrix
 converges to the period matrix $\Pi_\Sigma$ of the surface. Which is the
 claim of Th.~\ref{th:critic}.

 In the paper~\cite{CSMcC}, R.~Costa-Santos and B.~McCoy define a period
 matrix on a special cellular decomposition $\Gamma$ of a surface by squares.
 They don't consider the dual graph $\Gamma^{*}$.  Their period matrix is
 equal to one of the two diagonal blocks of the double period matrix we
 construct in this case. They don't have to consider the off-diagonal blocks
 because the problem is so symmetric that their period matrix is pure
 imaginary.

\subsection{Genus one} \label{sec:GenusOne}
Criticality solves partially the problem of having two different $g\times g$
period matrices instead of one since they converge to one another in a
refining sequence. However, on a genus one critical torus, the situation is
simpler: The overall curvature is null and a critical map is everywhere flat.
Therefore the cellular decomposition is the quotient of a periodic cellular
decomposition of the plane by two independant periods. They can be normalized
to $(1,\tau)$.  The continuous period matrix is the $1\times 1$-matrix
$\tau$. A basis of the two dimensional holomorphic $1$-forms is given by the
real and imaginary parts of $dZ$ on $\Gamma$ and $\Gamma^{*}$ respectively,
and the reverse. The discrete period matrix is the $2\times 2$ matrix $
\begin{pmatrix}
    \text{Im }\tau& \text{Re }\tau\\
    \text{Re }\tau& \text{Im }\tau
\end{pmatrix}
$ and the period matrices on the graph and on its dual are both equal to the
continuous one.

For illustration purposes, the whole construction, of a basis of
harmonic forms, then projected onto a basis of holomorphic forms,
yielding the period matrix, can be checked explicitely on the critical
maps of the genus $1$ torus decomposed by square or
triangular/hexagonal lattices:

Consider the critical square (rectangular) lattice decomposition of a
torus $\Diam=(\mathbb{Z}e^{i\,\theta}+\mathbb{Z}e^{-i\,\theta})/
(2p\,e^{i\,\theta}+2q\,e^{-i\,\theta})$, with horizontal parameter
$\rho=\tan\theta$ and vertical parameter its inverse. Its modulus is 
$\tau=\frac{q}{p}e^{2\,i\,\theta}$. The two dual
graphs $\Gamma$ and $\Gamma^{*}$ are isomorphic.  An explicit harmonic
form $\alpha^{\Gamma}_{1}$ is given by the constant $1/2p$ on
horizontal and downwards edges of the graph $\Gamma$ and $0$ on all
the other edges.  Its holonomies are $1$ and $0$ on the $p$, resp. 
$q$ cycles.  Considering $1/2q$ and the dual graph, we construct in
the same fashion $\alpha^{\Gamma}_{2}, \alpha^{\Gamma^{*}}_{1},
\alpha^{\Gamma^{*}}_{2}$.  The matrix of inner products is
\begin{equation}
    (\alpha_{k},\alpha_{\ell})_{k,\ell}
    =  \frac{1}{\sin 2\theta}
    \begin{pmatrix}
        \frac{q}{p} &&& \cos 2\theta  \\
        &\frac{q}{p} & \cos 2\theta   \\
    & \cos 2\theta&     \frac{p}{q}\\
    \cos 2\theta&&&     \frac{p}{q}
   \end{pmatrix}
    \label{eq:GammaMatrixSq}
\end{equation} 
using $\frac{\rho+1/\rho}{2}=1/\sin 2\theta$ and 
$\frac{\rho-1/\rho}{2}=-1/\tan 2\theta$ so that the period matrix is
\begin{equation}
    \Pi=\frac{q}{p}
    \begin{pmatrix}
        i\,\sin 2\theta&\cos 2\theta\\
        \cos 2\theta&i\,\sin 2\theta
    \end{pmatrix}.
    \label{eq:PeriodMatrixSq}
\end{equation}
Therefore there exists a holomorphic form which has the same periods
on the graph and on its dual, it is the average of the two half forms
of Eq.~\eqref{eq:zetaDual} and its periods are
$(1,\frac{q}{p}e^{2\,i\,\theta})$ along the $p$, resp.  $q$ cycles,
yielding the continuous modulus.  This holomorphic form is simply
the normalized fundamental form $\frac{d Z}{p e^{-i\,\theta}}$.
 
In the critical triangular/hexagonal lattice, we just point out to the
necessary check by concentrating on a tile of the torus, composed of
two triangles, pointing up and down respectively. We show that there 
exists an explicit holomorphic form which has the same shift on the 
graph and on its dual, along this tile.  Let $\rho_{-},
\rho_{\backslash}$ and $\rho_{/}$ the three parameters around a given
triangle.  Criticality occurs when $\rho_{-}\,\rho_{\backslash}+
\rho_{\backslash}\,\rho_{/}+ \rho_{/}\,\rho_{-}=1$.  The form which is
$1$ on the rightwards and South-West edges and $0$ elsewhere is
harmonic on the triangular lattice.  Its pure imaginary companion on
the dual hexagonal lattice exhibits a shift by $i\,\rho_{\backslash}$
in the horizontal direction and $i\,(\rho_{\backslash}+\rho_{-})$ in
the North-East direction along the tile.  Dually, on the hexagonal
lattice, the form which is $\rho_{\backslash}\,\rho_{-}$ along the
North-East and downwards edges and $1-\rho_{\backslash}\,\rho_{-}$
along the South-East edges, is a harmonic form. Its shift in the
horizontal direction is $1$, in the North-East direction $0$, and its
pure imaginary companion on the triangular lattice exhibits a shift by
$i\,\rho_{\backslash}$ in the horizontal direction and
$i\,(\rho_{\backslash}+\rho_{-})$ in the North-East direction along
the tile as before.  Hence their sum is a holomorphic form with equal
holonomies on the triangular and hexagonal graphs and the period
matrix it computes is the same as the continuous one.  This simply
amounts to pointing out that the fundamental form $dz$ can be explicitely
expressed in terms of the discrete conformal data.

\section{Bilinear relations} \label{sec:Bilin}
 
\begin{propo} \label{prop:bilin}
     Given a canonical dissection $\aleph$, for two closed
    forms $\theta,\theta'\in Z^{1}(\Diam)$,
    \begin{equation}
        \iint_{\Diam}\theta\wedge\theta' = \sum_{j=1}^{g}\left(
        \oint_{\aleph_{j}}\theta\oint_{\aleph_{j+g}}\!\!\!\!\theta'-
        \oint_{\aleph_{j+g}}\!\!\!\!\theta\,\oint_{\aleph_{j}}\theta' \right);
        \label{eq:iintThetaDiam}
    \end{equation}
    for two closed
   forms $\theta,\theta'\in Z^{1}(\Lambda)$,
   \begin{equation}
       \iint_{\Diam}\theta\wedge\theta' = \sum_{j=1}^{2g}\left(
       \oint_{\aleph_{j}^{\Lambda}}\theta
       \oint_{\aleph_{j+2g}^{\Lambda}}\!\!\!\!\theta'-
       \oint_{\aleph_{j+2g}^{\Lambda}}\!\!\!\!\theta\,
       \oint_{\aleph_{j}^{\Lambda}}\theta' \right).
       \label{eq:iintTheta}
   \end{equation}
\end{propo}
\begin{demo}{\ref{prop:bilin}}
  Each side is bilinear and depends only on the cohomology classes of the
  forms.  Decompose the forms onto the cohomology basis $(\alpha_{k})$.  On
  $\Lambda$, use Eq~\eqref{eq:alephIntersecAlpha} for the LHS and the duality
  property Eq.~\eqref{eq:dualBasis} for the RHS. On $\diamondsuit$, use their
  counterparts.
\end{demo}

Notice that for a harmonic form $\theta\in\mathcal{H}^{1}(\Lambda)$,
the form $*\theta$ is closed as well, therefore its norm is given by
\begin{equation}
    \theta\in\mathcal{H}^{1}(\Lambda)\implies 
    \norm{\theta}^{2}= \sum_{j=1}^{2g}\left(
    \oint_{\aleph_{j}}\theta\oint_{\aleph_{j+2g}}\!\!\!\!*\bar\theta-
    \oint_{\aleph_{j+2g}}\!\!\!\!\theta\,\oint_{\aleph_{j}}*\bar\theta \right).
    \label{eq:iintThetaHarmo}
\end{equation}

\section{Criticality and exponentials}\label{sec:criticality}
\subsection{Integration at criticality}
\label{sec:Integration}
We call a local map $Z:\,\Diam\supset U\to\mathbb{C}$ from a
connected, simply connected subcomplex to the euclidean plane,
\emph{critical} iff it is locally injective, orientation preserving
and the faces of $\Diam$ are mapped to rhombi such that the ratio of
the diagonal euclidean lengths is given by $\rho$.  The common length
$\delta$ of these rhombi is a characteristic of the map.  We showed
in~\cite{M,M01} that a converging sequence of discrete holomorphic
forms, on a refining (the lengths $\delta$ go to zero and faces don't
collapse) sequence of conformal maps of the same Riemann surface,
converge to a genuine holomorphic form.  The converse is also true,
every holomorphic form on a Riemann surface can be approximated by a
converging sequence of discrete holomorphic forms given a refining
sequence of critical maps.  And such a refining sequence exists for
every Riemann surface.

Of course, a very natural atlas of maps of a discrete Riemann surface is
given by the applications from every face, considered as a subcomplex,
to a well shaped rhombus in $\mathbb{C}$.  Such an application is only
well defined if two edges of the same face are not identified.  They
then form an atlas of the combinatorial surface and provide it with a
genuine structure of Riemann surface, namely a riemannian flat metric
with conic singularities~\cite{Tro}.  A discrete Riemann surface is
called \emph{critical} or critically flat if its conic singularities
are all multiple of $2\pi$.

We've seen that given a function $f$ and a $1$-form $dZ$, the residue
theorem implies that if they are both holomorphic then $f\,dZ$ is a closed
$1$-form.  The key point about criticality is that, given a critical
map $Z$ and a discrete holomorphic function $f$, one can construct the
\emph{holomorphic} $1$-form $f\,dZ$ 
\begin{equation}
    f\, dZ\,:\, \int_{(x,y)} f\, dZ = \frac{f(x)+f(y)}{2} 
    \left(Z(y)-Z(x)\right).
    \label{eq:fdZ}
\end{equation}
Reciprocally, every holomorphic $1$-form is locally of such a form, unique up
to $\varepsilon$.  The change of map acts as expected (see
Prop.~\eqref{prop:Leibnitz}) so for a critical discrete Riemann surface,
where the conic singularity angles are multiple of $2\pi$, $f\,dZ$ can be
continued as a well defined $1$-form on the universal covering of $\Diam$
wherever $f$ is defined.  Be careful that for another holomorphic form
$\theta\in\Omega(\Diam)$, the $1$-form $f\,\theta\in Z^{1}(\Diam)$ is only
\emph{closed} but not holomorphic in general.

\subsection{Exponential}\label{sec:Exp}
As an interesting example of discrete holomorphic function we present here
the discrete exponential of a constant argument. It is the first step in
trying to define the exponential map needed to provide Abel's correspondence
between divisors and meromorphic functions.

Different attempts have been made to define a discrete exponential,  
see~\cite{AB,Bob96} and references therein.

\begin{defi}
        Let $Z$ a critical map and $O\in\Lambda_{0}$ such that
        $Z(O)=0$, the origin.  For $\lambda\in\mathbb{C}$, define the
        holomorphic function \emph{exponential}
        $\text{Exp}({:}\lambda{:})\in\Omega^{0}(\Lambda)$, denoted
        $z\mapsto\text{Exp}({:}\lambda{:}\,z)$, by
        \begin{equation}
            \begin{cases}\text{Exp}({:}\lambda{:}\,O)=1\\ 
        d\text{Exp}({:}\lambda{:}\,z)=\lambda \text{Exp}({:}\lambda{:}\,z)dZ.
        \end{cases}
            \label{eq:ExpDef}
        \end{equation}
\end{defi}

\begin{propo}\label{prop:ExpWellDef}
If $|\lambda|\neq 2/\delta$, $\text{Exp}({:}\lambda{:})$ is a well defined
holomorphic function.
\end{propo}
\begin{demo}{\ref{prop:ExpWellDef}}
    If $(x,y)\in\Diam_{1}$ is an edge, Eq.\eqref{eq:ExpDef} reads
    \begin{equation}
        \Exp({:}\lambda{:}\,y)-\Exp({:}\lambda{:}\,x)=\lambda\,
        \frac{\Exp({:}\lambda{:}\,y)+\Exp({:}\lambda{:}\,x)}{2}\,
        \left(Z(y)-Z(x)\right)
        \label{eq:Expxy}
    \end{equation}
    so that 
    \begin{equation}
        \Exp({:}\lambda{:}\,y)=\frac{2+\lambda\left(Z(y)-Z(x)\right)}
        {2-\lambda\left(Z(y)-Z(x)\right)}\;\Exp({:}\lambda{:}\,x).
        \label{eq:Expy}
    \end{equation}
   As the map $Z$ is critical, on the face $(x,y,x',y')\in\Diam_{2}$, 
   $Z(y)-Z(x)=Z(x')-Z(y')$ as well as  $Z(y')-Z(x)=Z(x')-Z(y)$ so that 
   the product of the four terms along the face
   \begin{equation}
       \frac{2+\lambda(y-x)}
               {2-\lambda(y-x)}\,
       \frac{2+\lambda(x'-y)}
                {2-\lambda(x'-y)}\,
       \frac{2+\lambda(y'-x')}
               {2-\lambda(y'-x')}\,
       \frac{2+\lambda(x-y')}
                {2-\lambda(x-y')}
        =1
       \label{eq:ExpWellDef}
   \end{equation}
   where we wrote $Z(z)$ as $z$ for readability.
   So $\Exp({:}\lambda{:})$ is a well defined function on the simply 
   connected sub-complex $U\subset \Diam$, its value at a point is the 
   product of the contributions of each edge along a given path 
   connecting it to the origin, and the result doesn't depend on the 
   path. It is also holomorphic as
   \begin{eqnarray}    
       \frac{\Exp({:}\lambda{:}\,y')-\Exp({:}\lambda{:}\,y)}{\Exp({:}\lambda{:}\,x')-\Exp({:}\lambda{:}\,x)}
       &=&\frac{\frac{2+\lambda\left(y'-x\right)}
                {2-\lambda\left(y'-x\right)} -
           \frac{2+\lambda\left(y-x\right)}
                {2-\lambda\left(y-x\right)}}
          {\frac{2+\lambda\left(y-x\right)}
                {2-\lambda\left(y-x\right)} 
           \frac{2+\lambda\left(x'-y\right)}
                {2-\lambda\left(x'-y\right)}
                -1}
                \nonumber\\
        &=&\frac{y'-y}{x'-x}=i\rho(x,x').
       \label{eq:ExpHolo}
   \end{eqnarray}
\end{demo}

We note that it is logical to state $\varepsilon=\Exp({:}\infty{:})=\pm 1$
on $\Gamma$ and $\Gamma^{*}$ respectively. 

\begin{figure}[htbp]
\begin{center}\epsfig{file=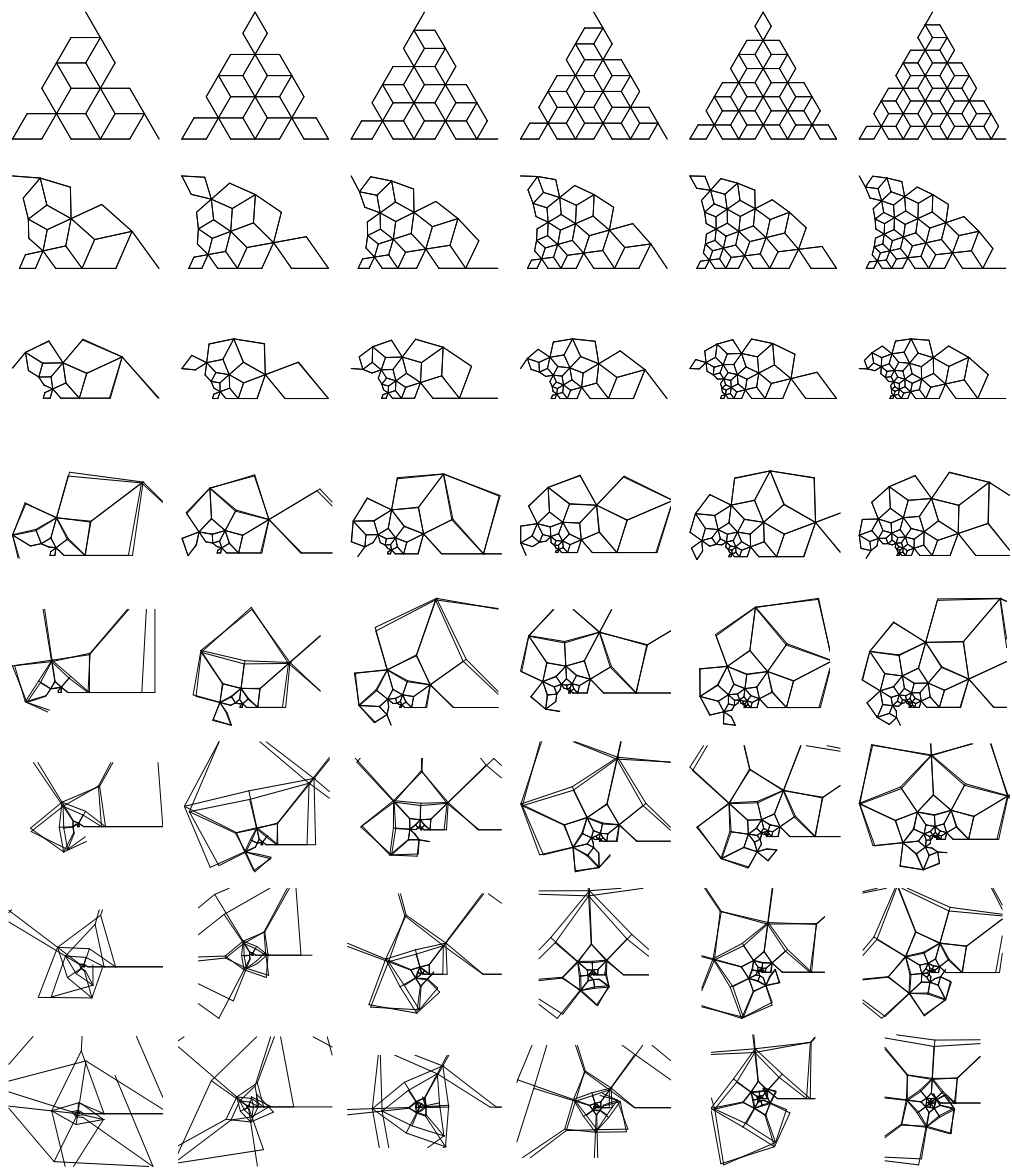}
\end{center}
\caption{The image of the first sextant of the triangular/hexagonal
lattice under the exponential, both discrete and continuous,
for growing parameters vertically and finer meshing horizontally.}
\label{fig:exp}
\end{figure}

On the rectangular lattice
$\Diam=\delta(e^{-i\theta}\mathbb{Z}+e^{i\theta}\mathbb{Z})$, the
exponential is given explicitly, for $z=\delta(n e^{-i\theta}+m
e^{i\theta})$ by
\begin{eqnarray}
    \Exp({:}\lambda{:}\,z)&=&
\left(\frac{1+\frac{\lambda\delta}{2}e^{i\theta}}
{1-\frac{\lambda\delta}{2}e^{i\theta}}\right)^{n}
\left(\frac{1+\frac{\lambda\delta}{2}e^{-i\theta}}
{1-\frac{\lambda\delta}{2}e^{-i\theta}}\right)^{m}
\notag\\
&=&\exp(\lambda z)
+O(\delta^{2})
    \label{eq:ExpRect}
\end{eqnarray}
as $\delta$ goes to zero keeping $z\in\mathbb{C}$ fixed, because
$(1+x)^{n}=\exp\left(n\log(1+x)\right)=\exp(n
x)+O\left((nx)^{2}\right)$.  Similar results are obtained for other
lattices, see Fig.~\ref{fig:exp} for the comparison of the discrete
and continuous exponentials on the triangular/hexagonal double.  The
discrepancy between the two increases with the modulus of the
parameter and decreases as the mesh is refined. 
 More generally, we will prove in~\cite{M0206041}  the
same behavior for any refining sequence of critical maps of a critical 
discrete Riemann surface.

\subsection{Change of base point}
Given another critical map $\zeta:V\to\mathbb{C}$ with an origin
$b\in\Lambda_{0}$, there exists a complex number $a$ such that
$\zeta=a(Z-b)$ on a connected component of the intersection $U\cap V$. 
If the discrete Riemann surface is itself critical and if $U\cup V$ is
contained in a ball then this number $a$ is the same for all connected
components.  As $d\zeta=a\, dZ$, it follows that
\begin{equation}
    \Exp_{\zeta}({:}\lambda{:}\,z)=\frac{\Exp_{Z}({:}a\,\lambda{:}\,z)}{\Exp_{Z}({:}a\,\lambda{:}\,b)}
    \label{eq:ExpMapChg}
\end{equation}
as expected.  Hence, on a critical discrete Riemann surface, the
exponential can be continued as a well defined holomorphic function on
the universal covering of the discrete surface. 

\subsection{Train-tracks}
For $\lambda$ with a norm equal to $2/\delta$ in a given critical map,
the associated exponential may not be defined as straightforwardly. 
We prove that in the compact case, in general it can only be defined
as zero:

We name the equivalence class of an (oriented) edge, generated by the
equivalence relation relating opposite sides of the same rhombi, a
\emph{train-track}, \emph{medial cycle} or \emph{thread} as it can be
seen graphically as a train-track succession of parallel edges of
rhombi, graph theoretically as a cycle in the medial
graph~\cite{M,GS87} or knot theoretically as a thread in the
projection of a link.

Given the train-track of an edge $(x,y)\in\Diam$ in a critical map $Z$, it
defines a complex number $\lambda=2/(Z(y)-Z(x))$, of norm $2/\delta$, which
verifies that $\Exp({:}\lambda{:})$ is null on one side of the train-track,
$\Exp({:}-\lambda{:})$ on the other.  If the sum of train-tracks parallel to
this one is non null in homology on the universal covering, then $\Exp({:}\pm
\lambda{:})\equiv 0$ everywhere. If it were null however,
$\Exp({:}\lambda{:})$ would be non trivial and defined independently on each
connected component.

\begin{propo}
    \label{prop:TrainTrack}
    On a critical discrete Riemann surface, a train-track is never
    null in homology.
\end{propo}
\begin{demo}{\ref{prop:TrainTrack}}
    Refining by splitting each quadrilateral in four doesn't change
    neither criticality nor the homology class of a thread, so we can
    suppose that if a thread is homotopic to zero then we have a
    finite domain which completely contains it.  Because all the conic
    angles are multiple of $2\pi$, this domain is a branched covering
    of a region of the euclidean complex plane.  The train-track is
    characterized by the conserved angle of each parallel side of its
    rhombi.  As all the angles of the rhombi are positive, the
    train-track follows a directed walk and can not backtrack to
    close.  Hence no train-track is null in homology.
\end{demo}

In the compact case, the fundamental domain is glued back into a
closed flat surface with conic singularities multiple of $2\pi$ hence
the fundamental group acts by a discrete subgroup $G$ of the Galilean
group of translations and rotations.  A corollary of the proposition
is that for all the angles appearing in this fundamental domain and
their image by $G$ (a discrete subset of the circle), the associated
$\Exp({:}\lambda{:})$ with $\lambda$ of modulus $2/\delta$ with these given
arguments must be defined as zero (or more precisely infinity).

We define a region $R\subset\Diam_{2}$ of a discrete Riemann surface
as \emph{convex} iff it is connected and for every pair of faces in
$R$ along the same thread, if the thread is open then every face in
between is also in $R$ and if it is closed, the same property holds
for at least one of the two halves of the thread.

Given a non convex set $D\subset \Diam_{2}$ contained in a disc, there
is a well defined procedure to find its minimal convex hull by adding
all the faces that are missing.  This ultimately reduces the number of
edges on the boundary as it doesn't introduce new threads by the
Jordan theorem so the procedure reduces the dimension of the space of
holomorphic functions, resp.  $1$-forms on it which are explicitly
$\partial D_{0}/2+1$, resp.  $\partial D_{0}/2-1$.  Notice that
$\partial D_{0}/2$ is the number of open threads segments contained in
$D$.  In the tensed case (see below), a basis of holomorphic functions
is provided by the set of functions which are non null on one side of
a given thread segment~\cite{CdV96} and $\varepsilon$.

\begin{conj}
    On a simply connected convex, the exponentials generate the space
    of holomorphic functions.
\end{conj}

What is clear is that on a non convex, they don't because in general,
not all the holomorphic functions defined on a non convex can be
continued into a holomorphic convex on the convex hull; however,
\begin{propo}
    \label{prop:convexContinuation}
In the simply connected case, every holomorphic function on a
connected non convex which can be continued into a holomorphic
function on the convex hull has a unique continuation; and for
every convex, there exists a maximal convex containing it where every
holomorphic function on the smaller convex can be uniquely analytically
continued.
\end{propo}

To prove this we need the notion of electrical moves.  We could also prove
Prop.~\ref{prop:TrainTrack} by implementing homotopy through
Reidemeister-like moves with spectral parameters.  They are called inversion
relation and star-triangle relation (linked to the Yang-Baxter equation) in
the context of integrable statistical models~\cite{Bax} or \emph{electrical
  moves} in the context of electrical networks~\cite{CdV96}.

\section{Electrical moves}\label{sec:electricalMoves}
\subsection{Definition}\label{sec:electricalMovesDef}
\begin{figure}[htbp]
\begin{center}\input{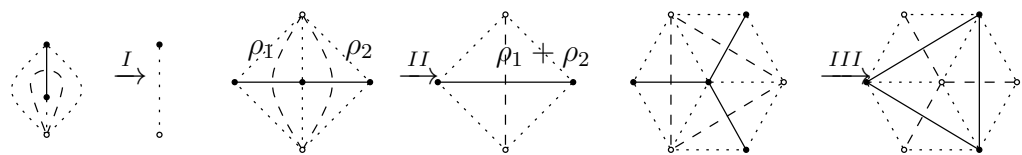}
\end{center}
\caption{The electrical moves.}    \label{fig:moves}
\end{figure}

The three types of electrical moves are presented in
Figure~\ref{fig:moves}.  They can be understood as moving, splitting
and merging the conic singularities around, keeping constant the
overall curvature, sum on the vertices of $2\pi$ minus the conic
angle.

The first move involves the elimination of a quadrilateral with any
discrete conformal parameter $\rho$ such that two of its adjacent
edges are identified.  Let's say $\rho$ labels the edge closed in a
loop, pictured as a dashed arc.  Before the move there are three
vertices, of angle $\theta$ at the black vertex, $\theta'$ at the
white vertex and $2\arctan 1/\rho$ at the cone summit.  After the
move, where the cone disappears, the angles are $\theta-2\arctan
1/\rho$ and $\theta'-4\arctan \rho$ respectively.  As $4\arctan
\rho+4\arctan 1/\rho=2\pi$, the total curvature has not changed.  In
statistical mechanics, this move amounts to normalizing the energy by
discarding autonomous self interaction.

The second move replaces two quadrilaterals, with conformal parameters
$\rho_{1}$ and $\rho_{2}$ along their respective parallel dashed
diagonals, identified along two adjacent edges, by a single
quadrilateral of conformal parameter the sum $\rho_{1}+\rho_{2}$. 
Dually, one replaces the series $\rho^{*}_{1}$ and $\rho^{*}_{2}$ by
$\frac{\rho^{*}_{1}\rho^{*}_{2}}{\rho^{*}_{1}+\rho^{*}_{2}}$.  These
formulae ring a bell to anyone versed in electrical networks as the
parallel and series replacement of two conductances by a single one. 
The conic angles change by
$2\arctan(\rho_{1}+\rho_{2})-2\arctan\rho_{1}-2\arctan\rho_{2})$ for
the white vertices and
$2\arctan(1/\rho_{i})-2\arctan\frac{1}{\rho_{1}+\rho_{2}}$ for each
black vertex.  It is called \emph{inversion relation} in statistical
mechanics.

The third move is the star-triangle transformation.  To three
quadrilaterals arranged in a hexagon, whose diagonals form a triangle
of conformal parameters $\rho_{1}$, $\rho_{2}$ and $\rho_{3}$, one
associates a configuration of three other quadrilaterals whose
diagonals form a three branched star with conformal parameters
$\rho'_{i}$ verifying
\begin{equation}
    \rho_{i}\rho'_{i}=\rho_{1}\rho_{2}+\rho_{2}\rho_{3}+\rho_{3}\rho_{1}
    =\frac{\rho'_{1}\,\rho'_{2}\,\rho'_{3}}{\rho'_{1}+\rho'_{2}+\rho'_{3}}.
    \label{eq:rho'rho}
\end{equation}
It is the \emph{star-triangle relation}, leading to the Yang-Baxter equation
in statistical mechanics.  Notice that if the central vertex is flat, then
the equation Eq.~\eqref{eq:rho'rho} is equal to $1$ and the conic angles at
either of the six external vertices don't change, hence flatness and
criticality are preserved by this move.  The relations for non flat angles
are more complicated and left to the reader.

Given two discrete Riemann surfaces related by
an electrical move, a holomorphic function or form defined on all the
shared simplices uniquely defines it on the simplices which are
different:

The value of a holomorphic function
\begin{itemize}
        \item  
at the conic singularity appearing in the type $I$ move has to be
equal to the value at the opposite vertex in the quadrilateral;
        \item  
at the middle point in the type $II$ move is the
$\left\{\frac{\rho_{2}}{\rho_{1}+\rho_{2}},
\frac{\rho_{1}}{\rho_{1}+\rho_{2}}\right\}$ weighted average of the
left and right values; 
        \item 
at the central point in the three pointed star configuration is the
$\left\{\frac{\rho_{i}}{\rho_{1}+\rho_{2}+\rho_{3}}\right\}$ weighted
average of the three boundary values.
\end{itemize}
  Moreover, the norm of the two related $1$-forms (or the coboundary
  of the related functions) are equal.  Hence the spaces of
  holomorphic forms for two electrically equivalent discrete Riemann
  surfaces are isomorphic, and isometric on $1$-forms.

The notion of convexity is stable by electrical move as well.

A discrete Riemann surface where no moves of type $I$ or $II$ are
possible is called \emph{tensed}~\cite{CdV96}.  On a disc,
electrically equivalent tensed discrete Riemann surfaces can be
obtained from one another through a series of type $III$ moves only
but it is no longer the case for non trivial topology.  Using the
Dirichlet theorem and the maximum principle, one shows~\cite{CdV96}
that tensed discrete Riemann surfaces have the property that for every
edge of the double graph $\Lambda$, there exists a holomorphic form
which is not zero on that edge.  It also provides with a basis of
holomorphic functions in the case of a topological disc, together with
$\varepsilon$, for each open thread, there is a holomorphic function
which is null on one side and nowhere null on the other, unique up to
a multiplicative constant~\cite{CdV96}.

\subsection{Demonstration of Prop.~\ref{prop:convexContinuation}}

\begin{demo}{\ref{prop:convexContinuation}}
    As electrical moves respect convexity and yield isomorphic
    holomorphic spaces, we can assume that the disc is tensed.  If
    there were several analytic continuations then it would contradict
    the fact that a basis of holomorphic functions is given by the set
    of functions which are non null on one side of a given thread
    segment.  The information contained in the value of a holomorphic
    function on a given convex allows to compute the value of the
    function on a greater convex. Wherever the value of the function is
    known at three vertices of a rhombi, it can be determined on the
    fourth, recursively closing the inwards corners.  The result is a
    convex set where each vertex on the boundary is adjacent to at
    least two faces which are not in the set.  On the square lattice
    $\Diam=\mathbb{Z}^{2}$ for example, knowing the value of a
    holomorphic function on the horizontal and vertical axis form the
    origin up to $(x,0)$ and $(0,y)$ allows to compute its values on
    the whole rectangle $\left((0,0),(x,0),(x,y),(0,y)\right)$.
\end{demo}
    
    The procedure of recursively closing the corners (and more
    elaborate conditions as well, implying more than three points)
    allows to try and extend a given holomorphic function defined on a
    non convex set but for every thread which is locally convexified,
    a condition for the values of the function has to be satisfied,
    if it is not satisfied then the function can be analytically
    continued into a \emph{meromorphic} functions in an essentially
    non unique way.

\subsection{Continuous moves}
These moves, while presented as discrete, are in fact continuous.  A
quadrilateral with a small or dually large conformal parameter $\rho$
is associated with a thin rhombus where two vertices are close to
one another hence the limit $\rho=0$ has to be understood as an absent
quadrilateral, where the edges on both sides are identified by pairs
and dually, an infinite conformal parameter along an edge means that
its two ends are identified into a single vertex.
\begin{figure}[htbp]
\begin{center}\input{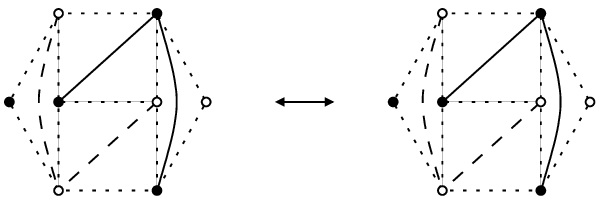}
\end{center}
\begin{eqnarray}
    \frac{\rho'_{D}}{\rho_{U}}&=& \frac{\rho_{D}}{\rho'_{U}}=
    \frac{\rho_{L}-\rho'_{L}}{\rho'_{R}-\rho_{R}}
    =(\rho_{L}-\rho'_{L})\frac{1+\rho_{U}\rho_{D}}{\rho_{U}}
    +\frac{\rho_{D}}{\rho_{U}}\notag\\ 
   &=&1/\left((\rho'_{R}-\rho_{R})\frac{1+\rho'_{U}\rho'_{D}}{\rho'_{D}}
        +\frac{\rho'_{U}}{\rho'_{D}}\right).
   \label{eq:movesCont}
\end{eqnarray}
\caption{The continuous electrical move.}    \label{fig:movesCont}
\end{figure}

Allowing for zero and infinite conformal parameters not only enables
to understand electrical moves as continuous, it also enables to
recover the first and second moves as special cases of the third. 
Notice that these moves do not preserve in general the complex
structure of the surface coded by the flat Riemannian metric
associated with the conic singularities.  However we will see that
criticality ensures that it is the case, the Riemann structure is 
unchanged by critical electrical moves.

\subsection{Reidemeister moves}
To recover isotopy and the usual Reidemeister moves, one has to allow
yet another type of discrete conformal structure, namely negative
parameters.  Geometrically speaking, a negative parameter corresponds
to a negative angle, or a change in the orientation, so the associated
quadrilateral takes part in a fold, a overhanging wrinkle in the
fabric of the surface.  Notice that the change of all the signs
corresponds to the complex conjugation at the level of functions or
forms: a holomorphic function on a graph with negative parameters is a
anti-holomorphic function on the same graph with positive parameters. 
We will call a graph with parameters of either signs a \emph{virtual}
discrete Riemann surface.  A holomorphic function on it can then be
understood as locally holomorphic, locally anti-holomorphic on a non
virtual discrete Riemann surface but it is not harmonic for this
positive discrete conformal structure so we won't make use of this
remark.  Notice that the total curvature, sum on all the vertices of
$2\pi$ minus the conic angle, is still unchanged by virtual electrical
moves.

%

Let's restrict ourselves to the case when all parameters on the graph
$\Gamma$ are equal or opposite, $\pm \rho$.  We associate to the
planar graph with signed edges $(\Gamma,\rho)$, a link in the tubular
neighborhood of the surface~\cite{M1}: The sign of each edge allows
the resolutions of the crossings occurring in the thread associated to
a given train-track into positive or negative regular projection
crossings.  There is of course an overall ambiguity of sign.  The
Reidemeister moves for this link are then recovered when we restrict
the second and third moves in a certain way:

We allow the second move only as the disappearance and appearance of a
pair of quadrilaterals of opposed conformal parameters $\rho, -\rho$,
leaving a collapsed quadrilateral of parameter zero.

In terms of catastrophe theory, the whole quadrilateral with negative
parameter is the unstable sheet of a cusp and its edges are the
bifurcation lines.  In this picture, the orientation of the surface is
only locally defined and changes across bifurcation lines where it is
not defined at all.

We restrict the third move to configurations where two parameters are 
equal and opposite to the third.

\begin{figure}[htbp]
\begin{center}\input{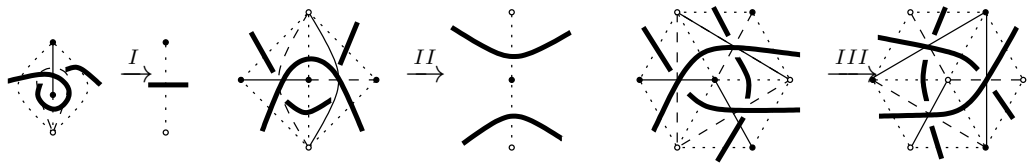}
\end{center}
\caption{The Reidemeister moves.}    \label{fig:movesReid}
\end{figure}

Notice that the usual non virtual discrete Riemann surfaces correspond
to alternating links.  Alternating links are special links so not all
the virtual discrete Riemann surfaces are Reidemeister equivalent to a
non virtual discrete Riemann surface.

Isotopy on the other hand is more free than the Reidemeister moves,
the signs and spectral parameters should be irrelevant.  But in fact,
on a critical map, we can keep track of them in a consistent way.  We
no longer restrict ourselves to maps where the parameters are all
equal or opposite and we free the restriction on the third move, but
we keep the restriction on the second move and forbid the first move. 
We call this set of allowed moves and the associated equivalence
relation \emph{critical isotopy}.  With these moves, isotopy can be
performed keeping criticality along the way.  These moves also have
the property to be stable by refining, two critical virtual discrete
conformal maps related by critical isotopy remain related if all their
quadrilaterals are cut into four (or more) smaller ones.  Hence,
refining to the continuous limit, the conformal structure of a class
of virtual discrete conformal maps (containing at least one genuine
discrete conformal map) is well defined.  Moreover, a given thread on
a map yields two homotopic threads on the refined map obtained by 
splitting a quadrilateral in four.

Notice that the first move can not occur on a critical map as it
involves a conic singularity of a given conic angle between $-\pi$
and $\pi$.  Notice as well that the sum of angles at a degree three
vertex, like the one appearing in a move of type $III$, can only be
$0$ or $\pm 2\pi$ on a virtual critical surface because it is the sum
of three angles between $-\pi$ and $\pi$ and has to be null mod $2\pi$
by criticality.

\section{Series} \label{sec:series}
As a first attempt to define divisors, and relate them to the degrees of
zeros and poles, we discuss here, after~\cite{Duf68}, discrete polynomials.

\subsection{Polynomials}\label{sec:Poly}
In a critical map $Z$, with $Z(O)=0$, we define recursively the powers
\begin{equation}
    Z^{k}:=\int_{O}^{z} k Z^{k-1}dZ
    \label{eq:ZkDef}
\end{equation}
with $Z^{0}=1$ and $Z^{1}=Z$.

Consider for example $\Diam$ containing the chain
$\{0,\frac{1}{n},\frac{2}{n},\ldots ,1\}$, the first $Z^{k}(x)$ with
$x=\frac{i}{n}$ are listed in Table~\ref{tbl:Zkx} and you show in
general, for a critical map of characteristic length $\delta$,
\begin{equation}
    |Z^{k}(x)-x^{k}|\leq \lambda_{k}|x|^{k-2}\delta^{2}
    \label{eq:Zkxk}
\end{equation}
with $\lambda_{k}$ independent of $x$ and $\delta$.  But this constant
is growing very rapidly with $k$, namely
$\lambda_{k}=\frac{k!}{2}\left(\frac{4}{\sin\eta}\right)^{k-2}$ where
$\eta$ is the smallest rhombus angle in the discrete Riemann surface.
\begin{table}[tbp]
    \centering
    \begin{equation*}
        \!\!\!\!\begin{array}{c|c|c|c|c|c|}
            k & 3 & 4 & 5 & 6 & 7  \\
            \hline&&&&&\\
            Z^{k}(x) & x^{3}\plus\frac{x}{2n^{2}} & x^{4}\plus\frac{2x^{2}}{n^{2}} & 
            x^{5}\plus\frac{5x^{3}}{n^{2}}\plus\frac{3x}{2n^{4}}& 
            x^{6}\plus\frac{10x^{4}}{n^{2}}\plus\frac{23x^{2}}{n^{4}} & 
            x^{7}\plus\frac{35x^{5}}{n^{2}}\plus\frac{49x^{3}}{n^{4}}\plus
            \frac{45x}{4n^{6}}  \end{array}
    \end{equation*}
    \caption{The first powers on the interval $[0,1]/n$ for $x=i/n$.}
    \label{tbl:Zkx}
\end{table}
It is for example not true that for a point close enough to the
origin $Z^{k}$ will tend to zero with growing $k$, on the contrary, if
$x$ is a neighbor of the origin with $(O,x)\in\Diam_{1}$ and $k\geq
1$, then $Z^{k}(x)=\frac{k!}{2^{k-1}} \,x^{k}$ in fact diverges with
$k$.  If $y$ is a next neighbor of the origin, with the rhombi
$(O,x,y,x')\in\Diam_{2}$ having a half angle $\theta$ at the origin,
$Z^{k}(y)=\frac{k!}{2^{2k-2}} \frac{\sin k\, \theta}
{\sin\theta \cos^{k-1}\theta} \,y^{k}$ has the same behavior and so
has every point at a finite distance of the origin.  It's only in the
scaling limit with the proper balance given by criticality that one
recovers the usual behavior $ |x|<1\implies 
|x^{k}|\xrightarrow[k\to\infty]{} 0$.

\begin{figure}[htbp]
\begin{center}\epsfig{file=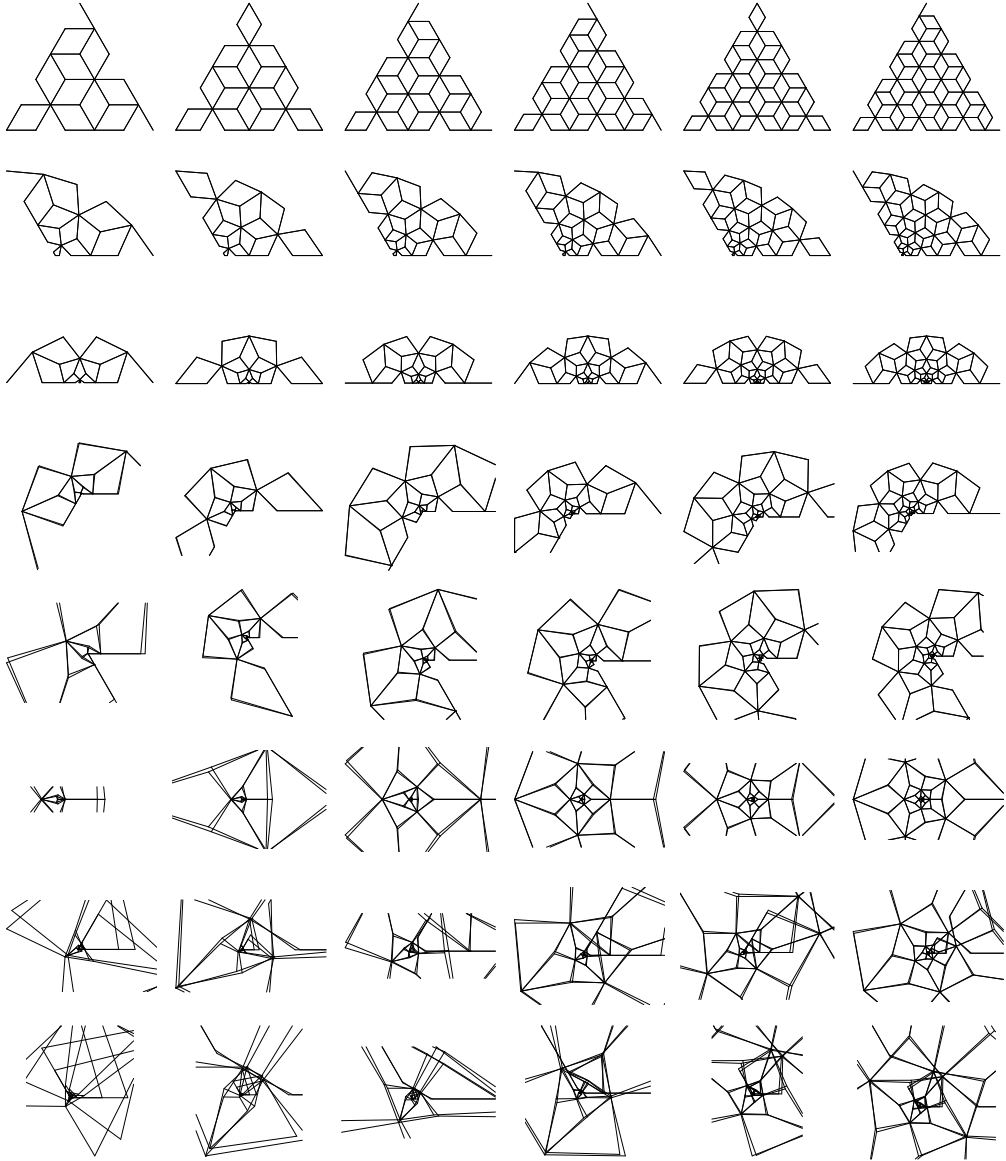}
\end{center}
\caption{The image of the first sextant of the triangular/hexagonal
lattice under the map $z\mapsto z^{k}$ both discrete and continuous,
for degree $1\leq k\leq 8$ vertically and finer meshing horizontally.}
\label{fig:Zk}
\end{figure}

See Fig.~\ref{fig:Zk} for the comparison of the discrete and
continuous powers on the triangular/hexagonal double.  The discrepancy
between the two increases with the degree (is null for degree up to
two) and decreases as the mesh is refined.

\subsection{Discrete Exponential as a series}\label{sec:ExpSeries}
The discrete exponential $\Exp({:}\lambda{:})$ is equal to the series
$\sum_{k=0}^{\infty}\frac{\lambda^{k}Z^{k}}{k!}$ whenever the latter is
defined as its derivatives fulfills the right equation and its value at the
origin is $1$.  The great difference with the continuous case is that the
series is absolutely convergent only for bounded parameters,
$|\lambda|<\frac{2}{\delta}$. This suggests that asking for a product such
that $\Exp({:}\lambda{:})\cdot\Exp({:}\mu{:})=\Exp({:}\lambda+\mu{:})$ may
not be the right choice.

\subsection{Ramification Number}\label{sec:Ramif}
Each rhombus $F=(x,y,x',y')\in\Diam$ is mapped by a holomorphic
function $f\in\Omega^{0}(\Lambda)$ to a quadrilateral
$(f(x),f(y),f(x'),f(y'))$ in the euclidean complex plane and generally
a cycle to a polygon.  If this polygon doesn't contain the origin 
$0\in\mathbb{C}$, we define
the \emph{ramification number} of $f$ at a cycle $\gamma$ to be its
winding number around $0$
\begin{equation}
    b_{f}(\gamma):=\oint_{f(\gamma)}\frac{dz}{z}.
    \label{eq:bfGamma}
\end{equation}
The ramification number of a face can only be $\{-1,0,+1\}$ because it 
is only four sided. If the ramification numbers at the faces inside an 
exact cycle $\gamma=\partial B$ are all defined, then the ramification 
of $f$ at $\gamma$ is the sum of these numbers. The ramification 
number of $Z$ at any loop is its winding number around $O$, twice for $Z^{2}$.
Unfortunately, things get more complicated for higher degrees.

\begin{conj}
    The ramification number of a polynomial of degree $k$ is at most 
    $k$, for all polynomial of degree $k$, there exists a cycle around 
    which it has ramification number $k$.
\end{conj}
It is true at the continuous limit, it can only change by a unit,
numerical evidence suggests that $Z^{k}$ has only non negative
ramification numbers but I don't know how to prove it in general.

\subsection{Holomorphic maps}\label{sec:HoloMap}
\begin{figure}[htbp]
\begin{center}\input{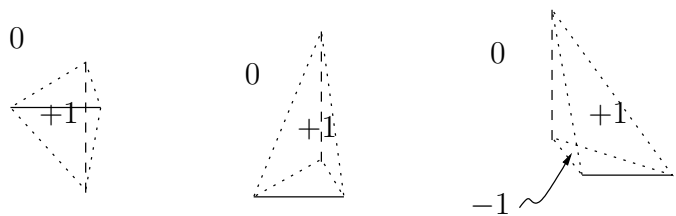}
\end{center}
\caption{The three shapes of quadrilaterals and their ramification 
number.}    \label{fig:notImmersed}
\end{figure}

A rhombus is mapped by a holomorphic function to a quadrilateral of
the plane which may or may not be immersed.  It is immersed whenever
the images of the diagonals whether cross or the line going through
one of them crosses the other (a kite), it is not immersed when the
line going through them cross at a point outside both of the segments. 
Because its two diagonals are mapped to vectors forming an orthogonal
direct basis, the only possibility for the ramification number to be
$-1$ is if the quadrilateral is not immersed and the origin is inside
the triangle winding the other way round
(see~Fig.\ref{fig:notImmersed}).  This case appears in the discrete
analog of $1/z$, in the derivative $\varepsilon'$, or in polynomials
of degree $2$ or more.

If we don't allow this situation to occur, we can define the notion of
holomorphic map between discrete Riemann surfaces.  We need a more
relaxed definition than criticality, namely
semi-criticality~\cite{M01} where rhombi are replaced by (not
necessarily convex) quadrilaterals. Note that this condition is 
sufficient to ensure the continuous limit theorem.

Let $(\Lambda,\rho)$ a discrete Riemann surface and $\Sigma$ the
associated topological surface.  It is \emph{semi-critical} for a flat
riemannian metric on $\Sigma$ iff the conic singularities are among
the vertices of $\Lambda$ and every face of $\Diam$ is realized by a
linear quadrilateral flattened so that its diagonals are orthogonal
and have a ratio of lengths governed by $\rho$.

If the quadrilateral is convex then its diagonals are well defined as
there exists a geodesic on the surface going from one of its four
vertices to the opposite one, staying inside the quadrilateral, a
usual linear segment.  When the quadrilateral is in the ``kite''
conformation however, one of the diagonals would be outside the flat
quadrilateral interior and its geometry may be disturbed by conic
singularities, it may not be a single segment anymore, it may even be
a single point, in particular the notion of orthogonality may not be
well defined: To compute an angle between two segments of geodesics,
they first have to be parallel transported so that they cross, turning
around a singularity alters this angle.  It is why the lengths and
orthogonality we are talking about are not those of geodesics on the
surface but those computed in the quadrilateral isometrically
flattened in the euclidean plane.

A discrete \emph{holomorphic map} $f$ between two semi-critical
discrete Riemann surfaces $(\Lambda,\rho)$, $(\Lambda',\rho')$ is a
cellular complex map, orientation preserving, locally injective
outside the vertices (where it can have a branching point) preserving
the conformal parameters.  It induces a genuine holomorphic map
between the underlying Riemann surfaces.  Reciprocally, a holomorphic
map to a Riemann surface associated with a semi-critical double
$(\Lambda,\rho)$ from another Riemann surface such that the branching
points are among the vertices of $\Lambda$ induces a semi-critical
double on the pre-image, hence a discrete holomorphic map.

In that case, the ramification number is well defined, so that with
$B:=\sum_{x\in\Lambda_{0}}b_{f}(x)$ the total ramification number,
$n=|f^{-1}(y)|$ for any $y$ in the image (counting multiplicities) the
degree of the map, $g$ and $g'$ the genuses of the two Riemann
surfaces, the Riemann-Hurwitz relation applies and
\begin{equation}
    g=n(g'-1)+1+B/2.
    \label{eq:RiemannHurwitz}
\end{equation}

This notion allows for the composition of discrete holomorphic maps.

\begin{propo}
    The exponential $\Exp({:}\lambda{:})$, for $|\lambda|<2/\delta$ in a
    critical map $Z: U\to\mathbb{C}$, is a holomorphic map from the
    discrete Riemann surface $U\cap\Diam_{2}$ to the semi-critical 
    discrete Riemann surface $Z(U)$.
\end{propo}
It suffices to check that every rhombus is mapped to a quadrilateral,
whether convex or of a kite form, it will imply local injectivity as
well.  One can assume that a given rhombus under consideration
contains the origin.  By inspecting the map
\begin{eqnarray}
    \mathbb{C} & \to & \mathbb{C}
    \nonumber  \\
    z & \mapsto & \frac{1+z}{1-z}
    \label{eq:MoebiusMap}
\end{eqnarray}
one sees that the quadrilateral $(1,\frac{1+ z'}{1- z'}, \frac{1+ z}{1-
z}\frac{1+ z'}{1- z'}, \frac{1+ z}{1- z})$ with $|z|=|z'|<1$ is always 
of that type, hence the result.

%

\subsection{Change of coordinate}\label{sec:Change}
We are now going to consider the change of coordinate for a critical map $Z$.
If $\zeta$ is another critical map with $\zeta=a\,(Z-b)$ on their common
definition set, the change of map for $Z^{k}$ is not as simple as the
Leibnitz rule $\left(a\,(z-b)\right)^{k}=a^{k}\sum_{j=0}^{k}\binom{k}{j}
z^{k-j}(-b)^{j}$ but is a deformation of it.  The problem is that pointwise
product is not respected, $Z^{k+\ell}(z)\neq Z^{k}(z)\times Z^{\ell}(z)$, in
particular the first is holomorphic and not the second, for each partition of
$k$ into a sum of integers, there is a corresponding monomial of degree $k$.
An easy inference shows that the result is still a polynomial in $Z$:
\begin{propo}\label{prop:Leibnitz}
    The powers of the translated critical map $\zeta=a(Z-b)$ are given by
    \begin{equation}
        \zeta^{k}=a^{k}\sum_{j=0}^{k}\binom{k}{j}(-1)^{j}Z^{k-j}B^{j}(b)
        \label{eq:yngSum}
    \end{equation}
    where  $B^{j}(b)$ corresponding to 
    $b^{j}$ is a  sum over all the degree $j$ monomials in 
    $b$, defined recursively by $B^{0}=1$ and
    \begin{equation}
    B^{k}(b):=\sum_{j=0}^{k-1}\binom{k}{j}(-1)^{k+j+1}Z^{k-j}(b)B^{j}(b)
        \label{eq:yngBk}
    \end{equation}
\end{propo}
The calculation is easier to read using Young diagrams, note the
product of pointwise multiplication monomial
\begin{equation}
    \left(Z^{k_{1}}(z)\right)^{\ell_{1}}
    \left(Z^{k_{2}}(z)\right)^{\ell_{2}} \ldots
    \left(Z^{k_{n}}(z)\right)^{\ell_{n}}
    \label{eq:yngZkl}
\end{equation}
with $k_{1}>k_{2}>\ldots >k_{n}$ as a Young diagram $Y$, coding
columnwise the partition of the integer given by the total degree
$k=\sum_{j=1}^{n}k_{j}\times\ell_{j}$ into the sum of
$\ell=\sum_{j=1}^{n}\ell_{j}$ integers.  For example the following
monomial of degree $15=3\times 2+2\times 4+1$ is noted \Yboxdim{.5em}
\begin{equation}
    \left(Z^{3}(z)\right)^{2}
    \left(Z^{2}(z)\right)^{4}  
    Z^{1}(z)=:\;\yng(7,6,2).
    \label{eq:yng}
\end{equation}
Then, $B^{j}(b)=\sum_{Y}c(Y)Y(b)$ where the sum is over all Young
diagrams of total degree $j$, $c(Y)$ is an integer coefficient that we
are going to define and $Y(b)$ is the pointwise product of the
monomials coded by $Y$ at $b$.  The coefficient of the Young diagram
$Y$ above is given by the multinomials
\begin{equation}
   c(Y)=(-1)^{k+\ell}\frac{k!}{(k_{1}!)^{\ell_{1}}(k_{2}!)^{\ell_{2}}\cdots 
   (k_{n}!)^{\ell_{n}}}\;\frac{\ell!}{\ell_{1}!\ell_{2}!\cdots \ell_{n}!}.
   \label{eq:yngCoeff}
\end{equation}

For example, the first Young diagrams have the coefficients
\Yboxdim{.5em} $c({\scriptsize\young(\hfil\hfil n\hfil\hfil)})=\frac{
n!}{1!\cdots 1!}\frac{n!}{n!}=n!$, $c({\scriptsize
\young(\hfil,\hfil,n,\hfil,\hfil)})=(-1)^{n+1}$,
$c({\scriptsize\young(\hfil\hfil,\hfil,n,\hfil,\hfil)})=
(-1)^{n+1}\frac{(n+1)!}{n!\,1!}\,\frac{2!}{1!\, 1!}=(-1)^{n+1}2{(n+1)}$
and $c({\scriptsize \young(\hfil\hfil
n\hfil\hfil,\hfil)})=-\frac{(n+1)!}{2!}\,\frac{n!}{1!\,(n-1)!}=
-\frac{(n+1)!\;n}{2}$.
The first few terms are listed explicitly in Table~\ref{tbl:Bk}. 

It
is to be noted that the formula doesn't involve the shape of the
graph, the integer coefficients for each partition are universal
constants and add up to $1$ in each degree.  As a consequence, since
$Z^{k}(z)Z^{\ell}(z)=Z^{k+\ell}(z)+O(1/\delta^{2})$, $k, \ell, z$
fixed, the usual Leibnitz rule is recovered in $O(1/\delta^{2})$. 
Let's stress again that these functions $B^{k}$ are discrete functions
on the graph $\Lambda$ which \emph{are not holomorphic}.  
\begin{table}[tbp]\Yboxdim{.5em}
    \centering
    \begin{eqnarray}
        B^{0} & = & +\;1
        \notag  \\
        B^{1} & = & +\;\yng(1)
        \nonumber  \\
        B^{2} & = & -\;\yng(1,1)+2\;\yng(2)
        \nonumber  \\
        B^{3} & = & +\;\yng(1,1,1)-6\;\yng(2,1)+6\;\yng(3)
        \nonumber  \\
        B^{4} & = & 
        -\;\yng(1,1,1,1)+8\;\yng(2,1,1)+6\;\yng(2,2)-36\;\yng(3,1)+24\;\yng(4)
        \nonumber  \\
        B^{5} & = & 
        +\;\yng(1,1,1,1,1)-10\;\yng(2,1,1,1)-20\;\yng(2,2,1)+60\;\yng(3,1,1)+
        90\;\yng(3,2)-240\;\yng(4,1)+120\;\yng(5)
        \nonumber  \\
        B^{6} & = & 
        -\;\yng(1,1,1,1,1,1)+12\;\yng(2,1,1,1,1)+30\;\yng(2,2,1,1)
        -90\;\yng(3,1,1,1)+20\;\yng(2,2,2)-360\;\yng(3,2,1)+480\;\yng(4,1,1)
        \nonumber\\
        &&-90\;\yng(3,3)+1080\;\yng(4,2)-1800\;\yng(5,1)+720\;\yng(6)
        \nonumber
    \end{eqnarray}
    \caption{The first analogs of $z^{k}$ needed in a change of basis.}
    \label{tbl:Bk}
\end{table}

The general change of basis of a given series however possible in
theory is nevertheless complicated and the information on the
convergence of the new series is difficult to obtain, even though
there are some exceptions like the exponential, if $\zeta=a(Z-b)$,
(see Eq.\eqref{eq:ExpMapChg}):
\begin{equation}
    \sum_{k=0}^{\infty}\frac{\lambda^{k}}{k!}\zeta^{k}\propto
    \sum_{k=0}^{\infty}\frac{(a\,\lambda)^{k}}{k!}Z^{k}.
    \label{eq:ExpMapChgSeries}
\end{equation}

\section{Conclusion}
The theory of discrete Riemann surfaces is shown to share a lot of
theorems and properties with the continuous theory, extending the
previous results to the matters related to the period matrix and
holonomies of holomorphic forms.  It begs for a good definition of the
order of a pole or a zero which would allow to prove the analog of
several crucial theorems of the continuous theory, namely Riemann-Roch
theorem, Abel's theorem and the Jacobi Inversion problem.  The main
challenge is to define a good discrete analog of the exponential of a
discrete holomorphic function so that the bilinear relations would
provide Abel's correspondences between divisors and meromorphic
functions.  The study of the discrete exponential $\Exp({:}\lambda{:}\, Z)$
for a critical map $Z$ and a constant $\lambda$ is a first step in
that direction.  Electrical moves are a crucial and powerful tool to
investigate the matter.  These combinatorial moves are clearly what is
needed to make the connection with Kasteleyn theory of pfaffians, on
the way to define discrete theta functions.

\section*{Acknowledgements}                     \label{sec:Acknowledgements}
I would like to thank B.~Mc~Coy and R.~Costa-Santos for asking me to publish
this material and for useful discussions and improvements. I thank Trevor
Welsh for simplifying Eq.~\eqref{eq:yngCoeff}.  This research is supported by
the Australian Research Council.

\bibliographystyle{unsrt} \bibliography{these}
\end{document}